\documentclass{emulateapj}

\def\msol{\hbox{\kern 0.20em $M_\odot$}}
\def\lsol{\hbox{\kern 0.20em $L_\odot$}}
\def\rsol{\hbox{\kern 0.20em $R_\odot$}}
\def\sr{\hbox{\kern 0.20em sr}}
\def\srmu{\hbox{\kern 0.20em sr$^{-1}$}}
 
\def\g{\hbox{\kern 0.20em g}}
\def\gmu{\hbox{\kern 0.20em g$^{-1}$}}
\def\kg{\hbox{\kern 0.20em kg}}
\def\pc{\hbox{\kern 0.20em pc}}
 
\def\mum{\hbox{\kern 0.20em $\mu$m}}
\def\mumd{\hbox{\kern 0.20em $\mu$m$^{-2}$}}
\def\cm{\hbox{\kern 0.20em cm}}
\def\m{\hbox{\kern 0.20em m}}
\def\km{\hbox{\kern 0.20em km}}
\def\nm{\hbox{\kern 0.20em nm}}
 
\def\s{\hbox{\kern 0.20em s}}
\def\h{\hbox{\kern 0.20em h}}
\def\sec{\hbox{\kern 0.20em sec}}
\def\min{\hbox {\kern 0.20em min}}
\def\smu{\hbox{\kern 0.20em s$^{-1}$}}
\def\smd{\hbox{\kern 0.20em s$^{-2}$}}
\def\an{\hbox{\kern 0.20em an}}
\def\anmu{\hbox{\kern 0.20em an$^{-1}$}}
\def\deg{\hbox{\kern 0.20em $^{\rm o}$}}
\def\yr{\hbox{\kern 0.20em yr}}
\def\yrmu{\hbox{\kern 0.20em yr$^{-1}$}}
\def\Myr{\hbox{\kern 0.20em Myr}}
\def\Mymu{\hbox{\kern 0.20em Myr$^{-1}$}}
\def\K{\hbox{\kern 0.20em K}}
\def\pcmu{\hbox{\kern 0.20em pc$^{-1}$}}
\def\pcmd{\hbox{\kern 0.20em pc$^{-2}$}}
\def\pcmt{\hbox{\kern 0.20em pc$^{-3}$}}
\def\kms{\hbox{\kern 0.20em km\kern 0.20em s$^{-1}$}}
\def\kmpd{\hbox{\kern 0.20em km$^{2}$}}
\def\kpc{\hbox{\kern 0.20em kpc}}
\def\cms{\hbox{\kern 0.20em cm\kern 0.20em s$^{-1}$}}
\def\erg{\hbox{\kern 0.20em erg}}
\def\ergs{\hbox{\kern 0.20em erg}}
\def\cmpd{\hbox{\kern 0.20em cm$^2$}}
\def\cmmd{\hbox{\kern 0.20em cm$^{-2}$}}
\def\cmms{\hbox{\kern 0.20em cm$^{-6}$}}
\def\cmpt{\hbox{\kern 0.20em cm$^3$}}
\def\cmmt{\hbox{\kern 0.20em cm$^{-3}$}}
\def\mpd{\hbox{\kern 0.20em m$^2$}}
\def\mmd{\hbox{\kern 0.20em m$^{-2}$}}
\def\mpt{\hbox{\kern 0.20em m$^3$}}
\def\mmt{\hbox{\kern 0.20em m$^{-3}$}}
\def\mujy{\hbox{\kern 0.20em $\mu$Jy}}
\def\mjy{\hbox{\kern 0.20em mJy}}
\def\Mj{\hbox{\kern 0.20em MJy}}
\def\jy{\hbox{\kern 0.20em Jy}}
\def\ghz{\hbox{\kern 0.20em GHz}}
\def\srmd{\hbox{\kern 0.20em sr$^{-1}$}}
\def \kms{km~$\rm{s}^{-1}$}

\def \mum{$\mu$m}
\def \gall{ {\it l} }
\def \galb{ {\it b} }

\def\G{\hbox{\kern 0.20em G}}

\def\h13cop{\hbox{H$^{13}$CO$^{+}$}}

\def\S+{\hbox{S{\small II}}}

\slugcomment{Accepted for publication in the Astronomical Journal}

\shorttitle{Galaxies Revealed in the ZoA}
\shortauthors{Marleau et al.}

\begin{document}

\newcommand{\jfourteen}{\hbox{$J=14\rightarrow 13$}} \title{Discovery
of Highly Obscured Galaxies in the Zone of Avoidance}

\author{F.R.\ Marleau\altaffilmark{1}, A.\
Noriega-Crespo\altaffilmark{1}, R.\ Paladini\altaffilmark{1}, D.\
Clancy\altaffilmark{1}, S.\ Carey\altaffilmark{1}, S.\
Shenoy\altaffilmark{1}, K.E.\ Kraemer\altaffilmark{2}, T.\
Kuchar\altaffilmark{3}, D.R.\ Mizuno\altaffilmark{3}, and S.\ Price\altaffilmark{2}}

\altaffiltext{1}{SPITZER Science Center, California Institute of
Technology, CA 91125.}
\altaffiltext{2}{Air Force Research Laboratory, Hanscom AFB, MA 01731.}
\altaffiltext{3}{Institute for Scientific Research, Boston College, 
Boston, MA 02135.}

\begin{abstract}
We report the discovery of twenty-five previously unknown galaxies
in the Zone of Avoidance. Our systematic search for extended
extra-galactic sources in the GLIMPSE and MIPSGAL mid-infrared surveys
of the Galactic plane has revealed two overdensities of these sources,
located around $l \sim 47$ and 55\arcdeg\ and $|b| \lesssim 1$\arcdeg\
in the Sagitta-Aquila region. These overdensities are consistent with
the local large-scale structure found at similar Galactic longitude
and extending from $|b| \sim 4$ to 40\arcdeg. We show that the infrared
spectral energy distribution of these sources is indeed consistent
with those of normal galaxies.  Photometric estimates of their
redshift indicate that the majority of these galaxies are found in the
redshift range $z \simeq 0.01 - 0.05$, with one source located at 
$z \simeq 0.07$. Comparison with known sources in the local Universe reveals 
that these galaxies are located at similar overdensities in redshift
space. These new galaxies are the first evidence of a bridge linking
the large-scale structure between both sides of the Galactic plane at
very low Galactic latitude and clearly demonstrate the feasibility of
detecting galaxies in the Zone of Avoidance using mid-to-far infrared
surveys.
\end{abstract}

\keywords{galaxies: distances and redshifts --- large-scale structure
of universe --- infrared: galaxies}

\lefthead{Marleau et al.}

\righthead{Galaxies Revealed in the ZoA}

\section{Introduction}

The last frontier in mapping the large-scale structure of the local
Universe is the Zone of Avoidance (ZoA). Starting in the late 1980s,
redshift surveys have provided an increasingly detailed picture of the
large-scale structure in the northern and southern Galactic
hemispheres, above and below the ZoA (e.g., CfA Redshift Survey,
Geller \& Huchra 1989; 2dF Redshift Survey, Colless 1999; Las Campanas
Redshift Survey, Shectman et al.\ 1996). However, the precise way the
coherent structures of the two halves connect remains largely unknown
and is not trivially predictable from the existing data. Moreover, it
is quite likely that undiscovered structures incorporating significant
mass concentrations lie behind the ZoA, as appears to be the case in
the region of the ``Great Attractor'' (Galactic longitude $l \sim
320$\arcdeg\ and latitude $b \sim 0$\arcdeg; Kolatt et al.\ 1995), and
finding these is essential for determining the dynamics of our local
Universe.

Many dedicated surveys of the ZoA have searched for hidden mass
concentrations of galaxies (see reviews by Kraan-Korteweg \& Lahav
2000; Kraan-Korteweg 2005). These have been primarily undertaken in
the optical (e.g., Roman, Iwata, \& Sait\=o 2000), near-infrared
(DENIS, Schr\"oder et al.\ 1999; 2MASS, Jarrett et al.\ 2000),
far-infrared (IRAS, Takata et al.\ 1996) and HI/radio (Parkes, see
e.g.\ Henning et al.\ 2000), resulting in a considerable reduction of
the ZoA. For example, the southern hemisphere HI survey with Parkes
was able to map over a thousand new galaxies within $|b| < 5$\arcdeg,
extending the prominence of the Norma supercluster (Radburn-Smith et
al.\ 2006). The most heavily investigated region, using the widest
wavelength regimes, has been the Great Attractor region, where a large
mass overdensity of 5$\times$10$^{16}$ solar masses has been predicted
from the systematic infall of 400 ellipticals (Dressler et al.\
1987). However, the most obscured regions ($|b| \leq 1$\arcdeg) of the
Milky Way, with visual extinction larger than 13~mag, remain largely
unexplored.

Recently, the {\em Spitzer Space Telescope} opened the door to further
advances and discoveries in the ZoA, with its significantly improved
detector technology, providing both better resolution and sensitivity
at mid- to far-infrared wavelengths (Werner et al.\ 2004). Indeed, one
of the first large surveys conducted by Spitzer was the Galactic
Legacy Infrared Mid-Plane Survey Extraordinaire (GLIMPSE, Benjamin et
al.\ 2003). GLIMPSE surveyed 2/3 of the inner Galactic disk with a
pixel resolution of 1.2\arcsec, using the Infrared Array Camera (IRAC)
at 3.6, 4.5, 5.8, and 8.0~\mum. The survey covered Galactic latitudes
$|b| < 1$\arcdeg\ and longitudes $|l| = 10 - 65$\arcdeg, both sides of
the Galactic center. Source confusion and saturation from the bright
star-forming regions forced the GLIMPSE team to adopt a relatively
conservative approach, mapping the Plane with a 4~second integration
time. In comparison, the Galactic First Look Survey (FLS, Burgdorf et
al.\ 2005) used 48~second exposures at $l = 97$\arcdeg\ and $|b| <
4$\arcdeg, to study the structure of the Galactic disk. The following
year, the GLIMPSE survey was complemented by the 24 and 70 Micron
Survey of the Inner Galactic Disk (MIPSGAL, Carey et al.\ 2008),
essentially covering the same region of the Plane at longer
wavelengths. One of the first extragalactic results that followed from
these dust-penetrating surveys, was the discovery of galaxies peering
through the most obscured portion of the Great Attractor (Jarrett et
al.\ 2007), demonstrating the discovery potential of {\em Spitzer} in
the ZoA.

\section{Data}

In this paper, we present the results of a galaxy search in the ZoA
based on data obtained from the two {\it Spitzer} Legacy Surveys of
the Galactic Plane; the Galactic Legacy Infrared Mid-Plane Survey
Extraordinaire (GLIMPSE) (Benjamin et al.\ 2003) and the MIPS Galactic
Plane Survey (MIPSGAL) (Carey et al.\ 2008).  These surveys cover
similar region, i.e.\ Galactic latitudes $|l| = 10\arcdeg - 65\arcdeg$
and one degree longitude above and below the plane ($|b| \lesssim
1$\arcdeg). 

The enhanced products from GLIMPSE were downloaded directly from the
Spitzer Science Center popular products
website\footnote{http://data.spitzer.caltech.edu/popular/glimpse/},
and included post-processed mosaics of the four IRAC (Fazio et al.\
2004) bands (3.6, 4.5, 5.8 \& 8\mum), as well as a detailed
description on how the data were reduced (Meade et al.\ 2007ab). The
GLIMPSE data were taken using two 2 sec exposures in order to deal with
some of the brightest regions in the Galactic plane, and therefore the
images remained shallow in the darkest regions. The quoted surface
brightness sensitivities (5$\sigma$) were 2.7, 2.2, 6.3, and 5.6 mJy
sr$^{-1}$ at 3.6, 4.5, 6.0 \& 8\mum, respectively. The angular
resolution of the enhanced mosaics were $\sim 2.5\arcsec$, with a
plate size of one square degree.

The MIPSGAL data were obtained using MIPS (Rieke et al.\ 2004) in its
fast scan mode. The mapping at 24\mum~relied on two passes, for a
minimum coverage of 10 samples per position, with a total integration
time of 30 sec, and with an extended source sensitivity of $\sim 0.16$
MJy sr$^{-1}$ (1$\sigma$). This scanning strategy was chosen so as to
compensate for the latent effects produced by bright sources. Due to
the smaller field of view (FOV) of the 70~\mum\ array compared to that
at 24~\mum\ ($\sim 2.\arcmin5 \times 5\arcmin$ versus $5\arcmin \times
5\arcmin$), the data were obtained using an interlaced pattern with a
minimum coverage of 5 samples per pixel, yielding a total integration
time of 15 sec and a sensitivity of $\sim 0.74$ MJy sr$^{-1}$
(1$\sigma$). The enhanced products, that will soon be delivered and
released (Carey et al.\ 2008), included post-processed mosaics at 24
and 70~\mum, with an angular resolution of $\sim 6$\arcsec\ and
20\arcsec\, respectively, and a plate size of 1.21 square degree 
($1.1\arcdeg \times 1.1\arcdeg$).

The IRAC and MIPS data were supplemented with near-infrared J, H and
Ks (1.25, 1.65 \& 2.17~\mum, respectively) data from the Two Micron
All Sky Survey (2MASS) (Skrutskie et al.\ 2006), downloaded directly
from the NASA/IPAC Infrared Science Archive
(IRSA)\footnote{http://irsa.ipac.caltech.edu/Missions/2mass.html}.

The boundaries of our search were set by the available processed
MIPSGAL data and the structure of the Galactic Plane itself (avoiding
very bright and confused regions), leaving us with approximately 50
square degrees (40\arcdeg $\ge$ l $\ge$ 65\arcdeg; -1 $\ge$ b $\ge$ 1)
to survey during our initial search. Our visual search for extended
sources in the 24~\mum\ MIPSGAL data revealed the first galaxy
candidate. In continuing our search, we found that using the GLIMPSE
enhanced product, with higher angular resolution, was a better way to
identify potential galaxy candidates. We made use of the red color (in
the composite IRAC 3.6, 4.5 and 8~\mum\ images) of these galaxies to
narrow down our search in the GLIMPSE plates. 

\section{Detection of Galaxies and Cross-Identification}

\begin{figure}[h!]
\centerline{\hbox{
\includegraphics[width=260pt,height=40pt,angle=0]{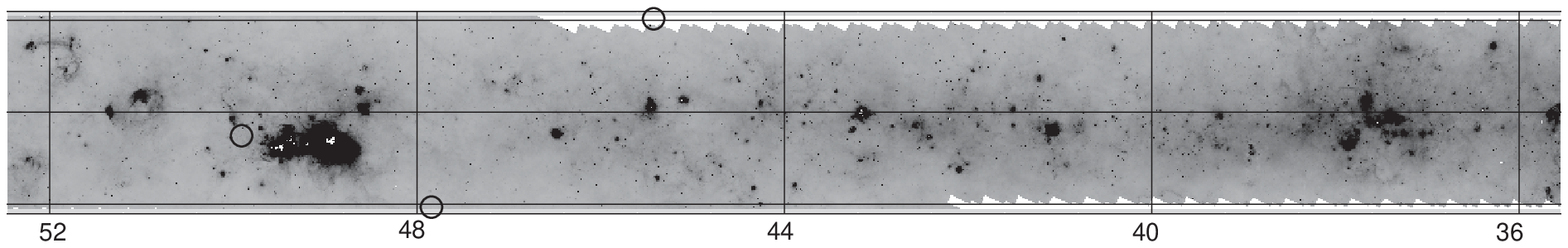}
}}
\centerline{\hbox{
\includegraphics[width=260pt,height=40pt,angle=0]{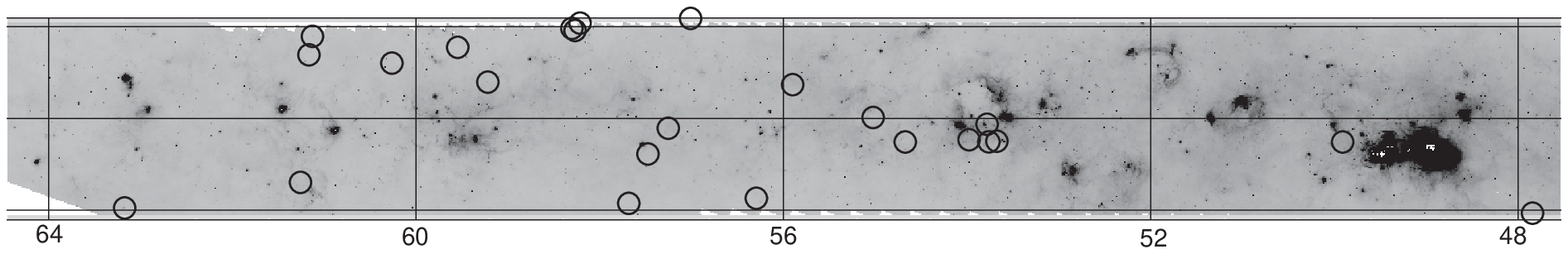}
}}
\caption{\label{fig:MIPSGAL} Location of the twenty-five newly
discovered galaxies, superposed on the 24~\mum\ MIPSGAL image
(5\arcsec\ per pixel) of the Galactic plane. The {\em top} panel
covers $l \sim 36 - 52\arcdeg$ and the {\em bottom} one $l \sim 48 -
64\arcdeg$, with $|b| \lesssim 1$\arcdeg.}
\end{figure}

\begin{figure}[h!]
\centerline{\hbox{
\includegraphics[width=220pt,height=220pt,angle=0]{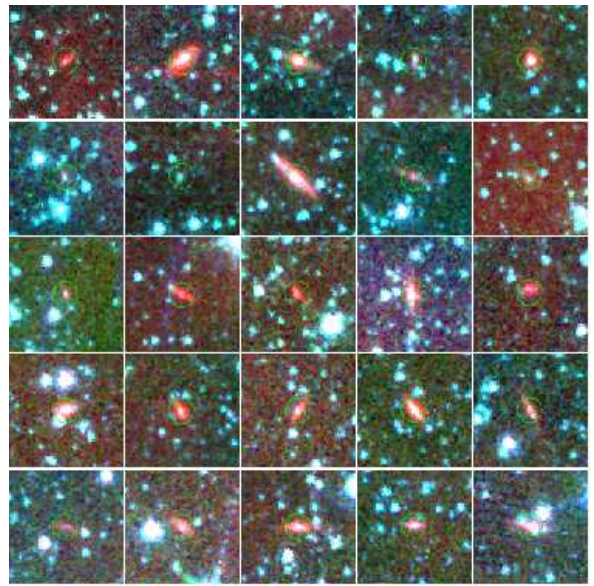}
}}
\caption{\label{fig:8um} IRAC 3.5 ({\em blue}), 4.5 ({\em green}), and
8~\mum\ ({\em red}) color composite postage stamps of our sample of
twenty-five newly discovered galaxies. The galaxies are ordered from
{\em top} to {\em bottom} and {\em left} to {\em right} following the
listing order in Table~\ref{tbl:coo}. Most of these galaxies are at
the edge of the GLIMPSE survey detection limit. The FOV of each image
is 53\arcsec\ $\times$ 53\arcsec\ (0.6\arcsec\ per pixel) with N up
and E to the left. The {\em green} circles overlayed on the images are
12\arcsec\ diameter in size. The scale parameters of the postage
stamps were optimized for each individual source and therefore differ
for each image.}
\end{figure}

Our systematic search, covering $\sim 25$\arcdeg\ in Galactic
longitude, successfully revealed two overdensities of a total of
twenty-five new galaxies, located around $l \sim 47$ and 55\arcdeg\ in
the Sagitta-Aquila region (see Table~\ref{tbl:coo} and
Figure~\ref{fig:MIPSGAL}). Their IRAC 3.5, 4.5, and 8~\mum\ color
composite images are shown in Figure~\ref{fig:8um}. Of these,
twenty-two were found to have a 24~\mum\ counterpart in the MIPSGAL
images, as three fell outside the MIPS survey, and nine had measurable
flux densities at 70~\mum.

\begin{figure}[h!]
\centerline{\hbox{
\includegraphics[width=220pt,height=220pt,angle=0]{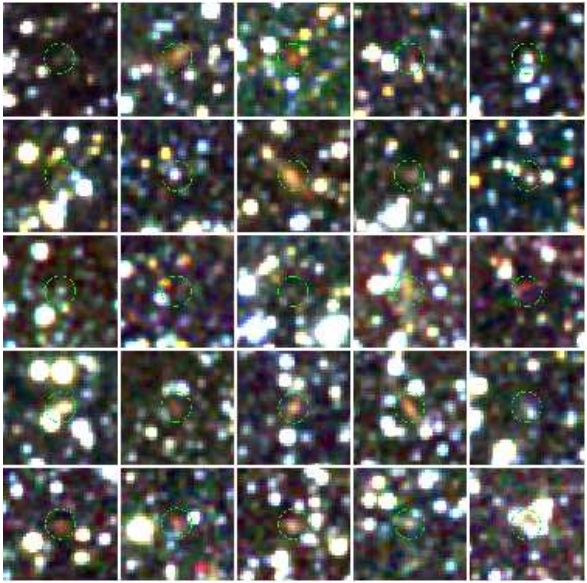}
}}
\caption{\label{fig:2mass} The 2MASS J ({\em blue}), H ({\em green}),
and Ks ({\em red}) color composite images of the same galaxies. The
galaxies are displayed in the same order as shown in
Figure~\ref{fig:8um}. These extended sources are barely visible in the
near-infrared where the extinction is higher, as seen in these
images. The FOV of each image is 43\arcsec\ $\times$ 43\arcsec\
(0.99972\arcsec\ per pixel) with N up and E to the left. The {\em
green} circles overlayed on the images are 12\arcsec\ diameter in
size. The scale parameters of the postage stamps were optimized for
each individual source and therefore differ for each image.}
\end{figure}

We cross-identified our candidate galaxies with their Two Micron All
Sky Survey (2MASS) J, H, and Ks counterparts (see
Figure~\ref{fig:2mass}). As can be seen in Figure~\ref{fig:8um} and
\ref{fig:2mass}, our galaxies are typically very faint in all IRAC
bands (median value of IRAC 8~\mum~=~8.9~mag), and just barely
visible in the 2MASS near-infrared images (median value of 
Ks 2.17~\mum~=~11.9~mag).

\begin{figure}[h!]
\centerline{\hbox{
\includegraphics[width=220pt,height=220pt,angle=0]{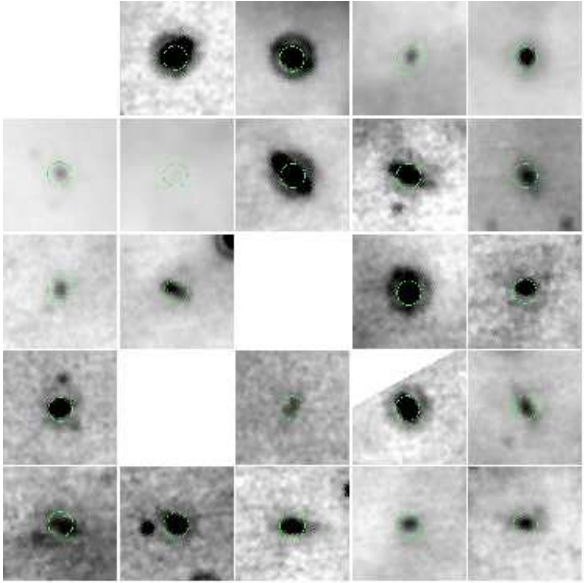}
}}
\caption{\label{fig:24um} MIPS 24~\mum\ postage stamps of the same
galaxies. The galaxies are displayed in the same order as shown in
Figure~\ref{fig:8um}. The candidate galaxies are essentially point
sources at 24~\mum. The FOV of each image is 54\arcsec\ $\times$
54\arcsec\ (1.25\arcsec\ per pixel) with N up and E to the left. The
{\em green} circles overlayed on the images are 12\arcsec\ diameter in
size. The images are displayed in reverse colormap and the scale
parameters of the postage stamps were optimized for each individual
source and therefore differ for each image.}
\end{figure}

\section{Photometric Measurements}

The spectral energy distribution (SED) of each candidate galaxy was
derived using photometric measurements obtained in all nine bands
(from 1.25 to 70~\mum). The aperture position and size for each object
were fixed in order to measure the colors or broadband SED at the same
physical region of each galaxy. The photometry was done using an
aperture radius of 6\arcsec\ and the background contributing to the
total flux density within the same aperture was removed. This aperture
size was chosen to avoid as much as possible contamination due to
foreground stars, mostly seen in the near-infrared images, while
matching the average size, i.e.\ two times the full width at half
maximum (FWHM), of the point spread function (PSF) at 24~\mum\ (FWHM
of 6\arcsec) given that all candidate galaxies are essentially point
sources at 24~\mum\ (see Figure~\ref{fig:24um}). Nevertheless, the
flux density measurements of some galaxies, e.g.\
SPITZER193633+225125, had to be corrected due to contamination from
foreground stars. This was done by interpolating over the neighboring
galaxy pixel values.

For the 2MASS measurements, a zero point was estimated using a point
source in the field with photometric measurements provided by the
2MASS point source catalog\footnote{http://irsa.ipac.caltech.edu/}. No
aperture correction was applied as it is essentially
negligible\footnote{see
http://www.ipac.caltech.edu/2mass/releases/allsky/doc/sec4\_4c.html}.
Dust extinction towards each galaxy was inferred from the IRAS/DIRBE
far-infrared maps (Schlegel et al.\ 1998) (see
Table~\ref{tbl:ext}). The values of A(V) at the reference point were
transformed to the 2MASS passbands using the extinction curve of Dutra
et al.\ (2002), where A(J)~=~0.301~A(V), A(H)~=~0.180~A(V), and 
A(Ks)~=~0.118~A(V). The values of A(Ks) for the candidate galaxies range
from 0.9 to 3.7~mag.

The IRAC flux densities were multiplied by the aperture correction
factor according to the formulation derived by the IRAC
team\footnote{http://ssc.spitzer.caltech.edu/irac/calib/extcal/}. The
aperture correction for an extended source and for a 6 arcsec radius
aperture was 1.02778 (channel 1), 1.07868 (channel 2), 1.00985
(channel 3) and 0.965037 (channel 4). The zero magnitude flux
densities were taken from Reach et al.\ (2005) to be 280.9 (channel
1), 179.7 (channel 2), 115.0 (channel 3), and 64.13 Jy (channel
4). Finally, the extinction corrections at these wavelengths were
obtained using the transformation given in Indebetouw et al.\ (2005)
(their equation 4; see Table~\ref{tbl:ext} for computed values). The
aperture correction was applied to the MIPS flux densities following
the MIPS data handbook prescription (Engelbracht et al.\ 2007; Gordon
et al.\ 2007). The aperture correction for a 6 arcsec radius aperture
was 1.7 at 24~\mum\ and 3.9 at 70~\mum. The extinction corrections at
these wavelengths were derived using the extinction curve of
Indebetouw et al.\ (2005) and are listed in Table~\ref{tbl:ext}.

All extinction-corrected photometric measurements are given in
Table~\ref{tbl:phot}. Uncertainties in the flux density measurements
varied with passband. In the shorter 2MASS wavelengths, the
photometric uncertainties were dominated by the extinction
correction. The upper and lower error bars on the 2MASS flux density
measurements were computed using the maximum and minimum values of
A(V) from Schlegel et al.\ (1998) derived using a five arcminute
radius circle for averaging centered on the reference position. For
the Spitzer bands, we assumed flux density uncertainties of the order
of 20\%.

\section{Spectral Energy Distribution}

\begin{figure}[h!]
\centerline{\hbox{
\includegraphics[width=130pt,height=130pt,angle=0]{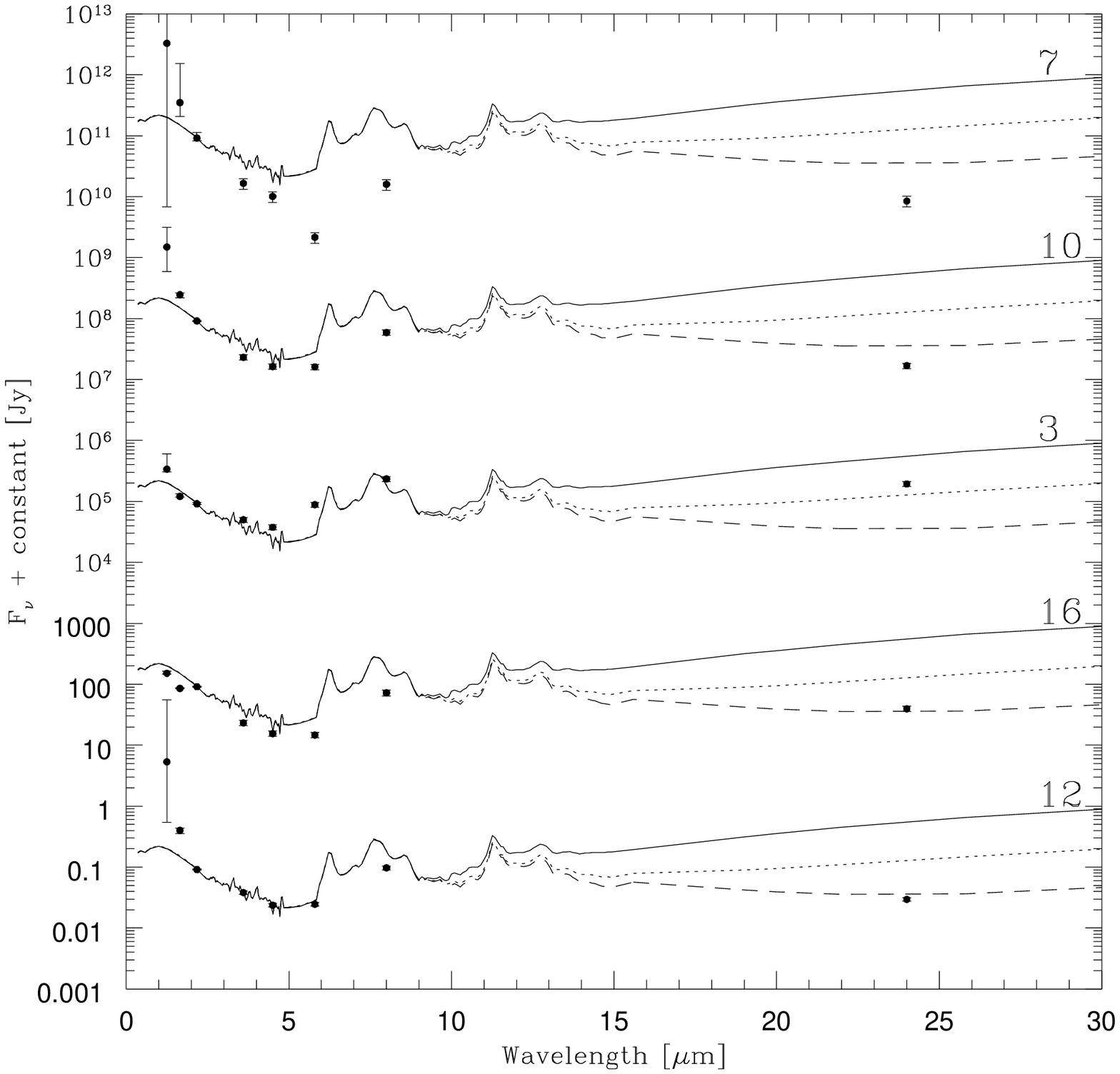}
\includegraphics[width=130pt,height=130pt,angle=0]{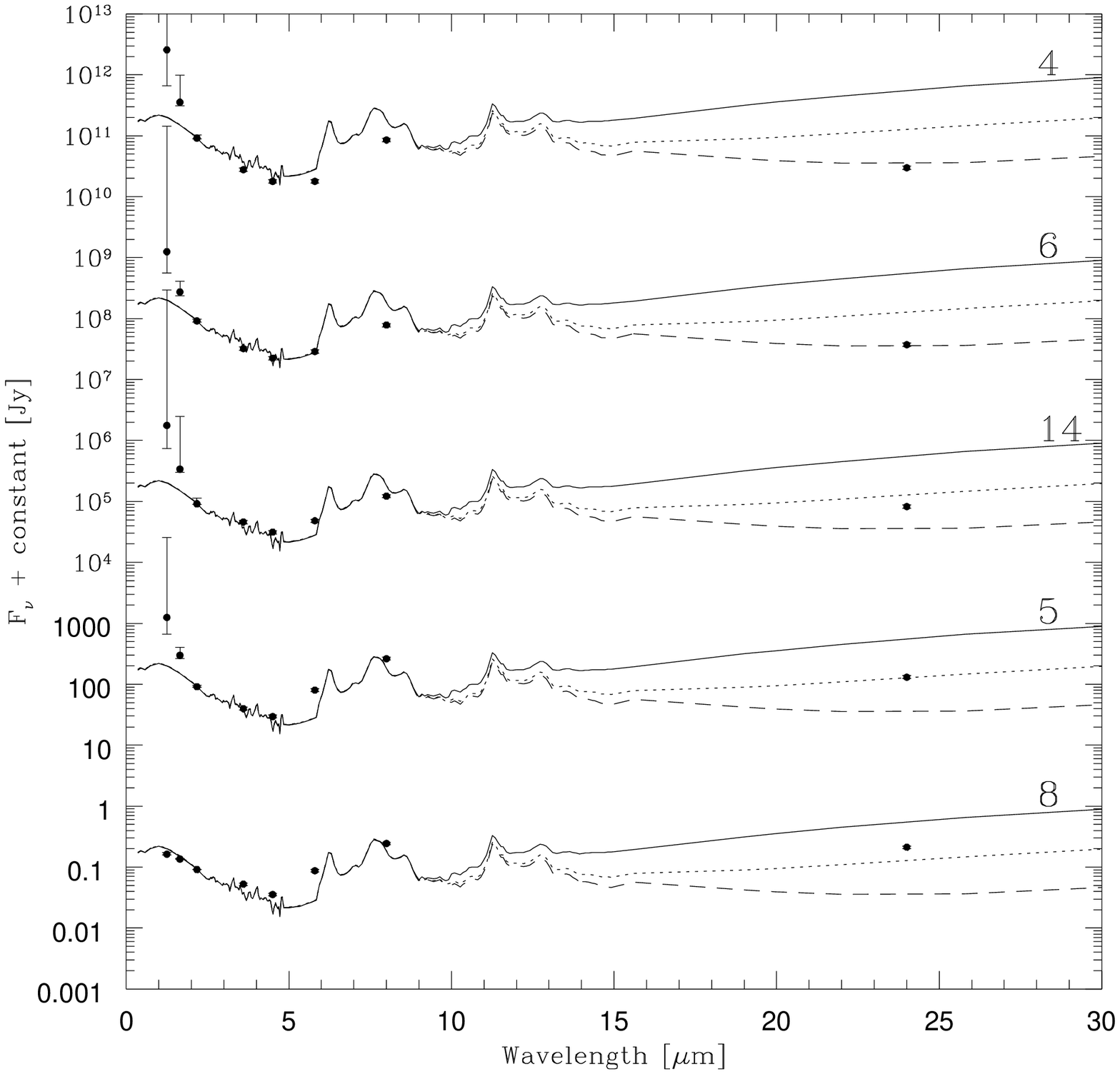}
}}
\centerline{\hbox{
\includegraphics[width=130pt,height=130pt,angle=0]{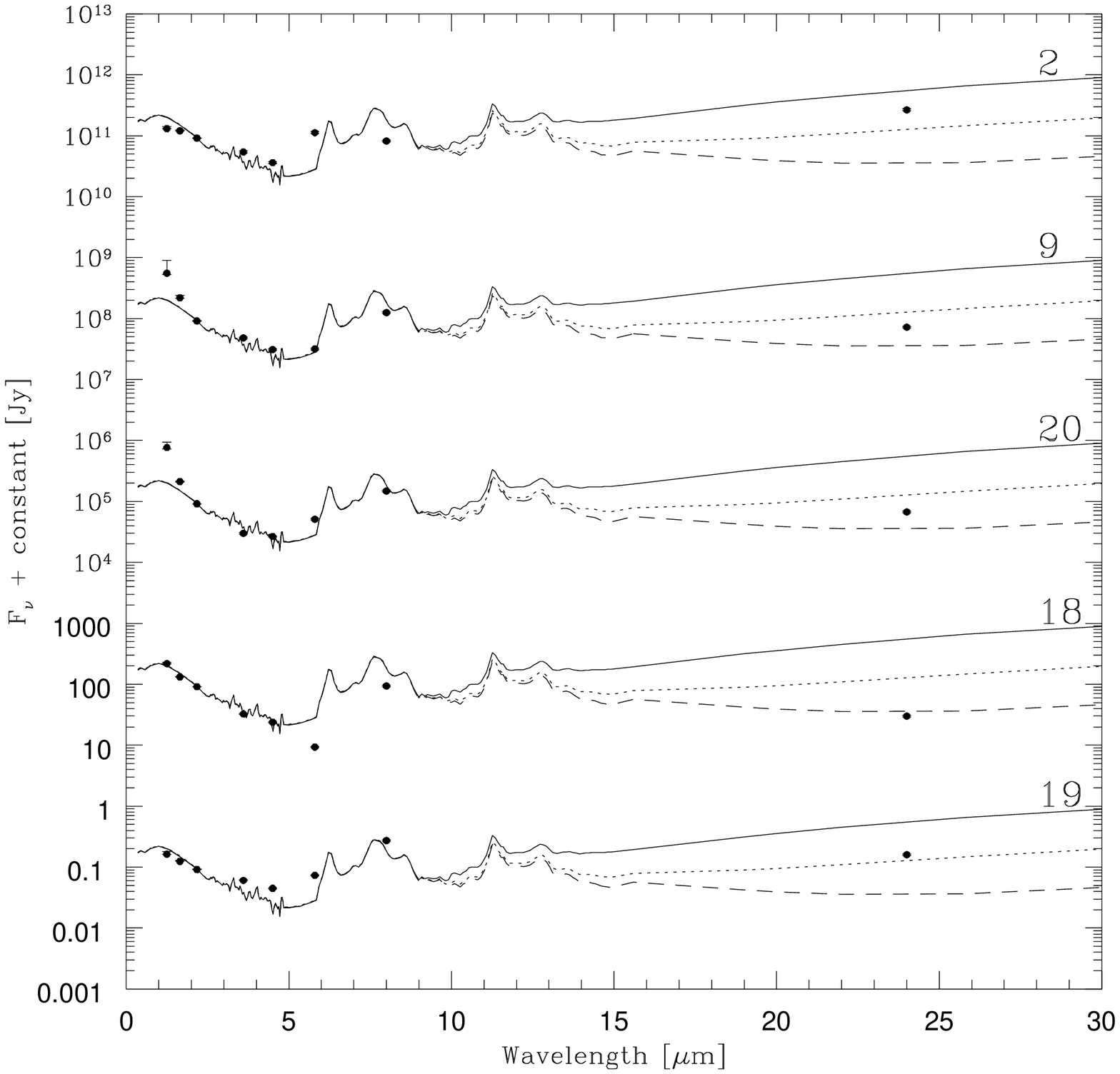}
\includegraphics[width=130pt,height=130pt,angle=0]{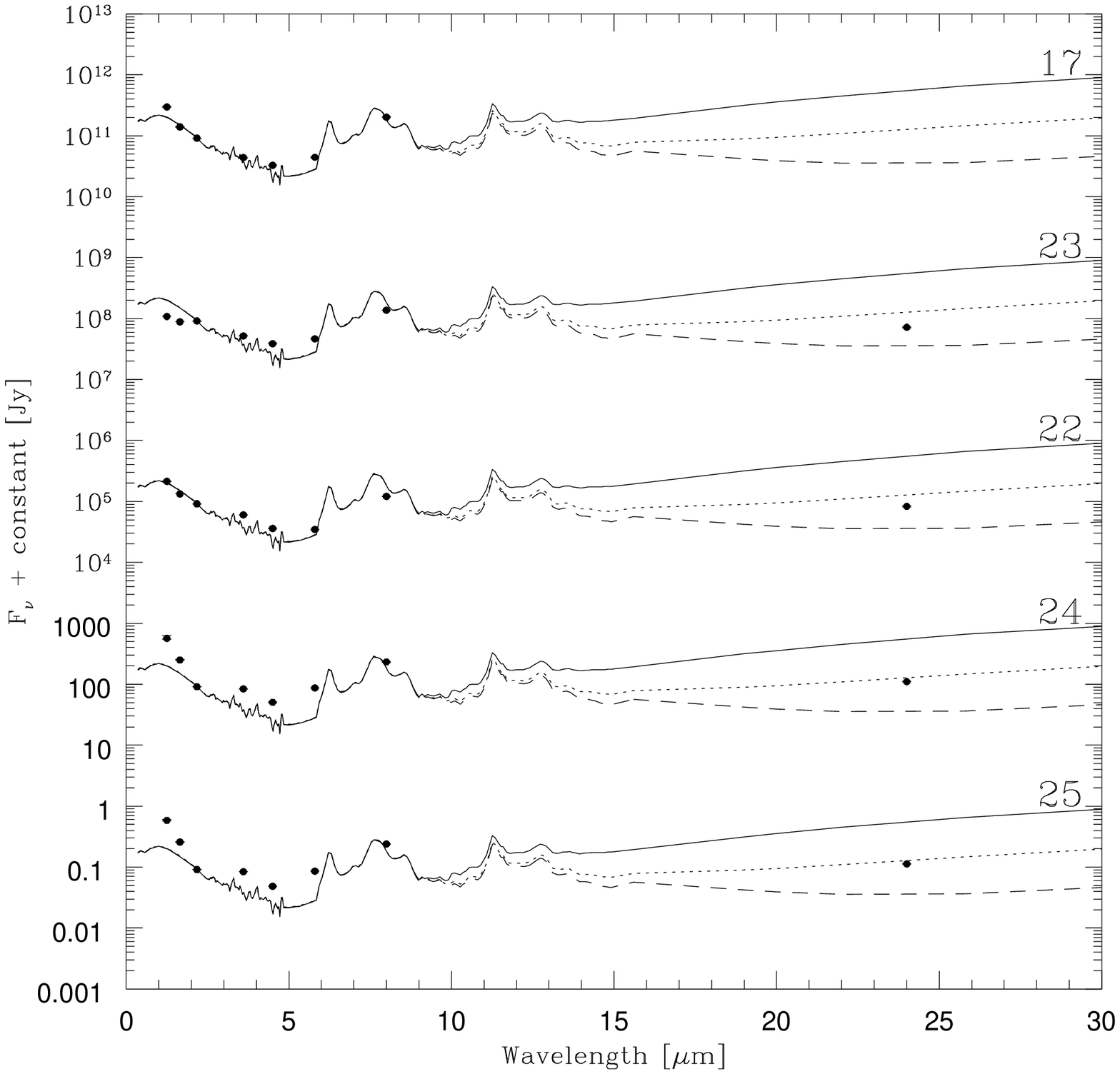}
}}
\centerline{\hbox{
\includegraphics[width=130pt,height=130pt,angle=0]{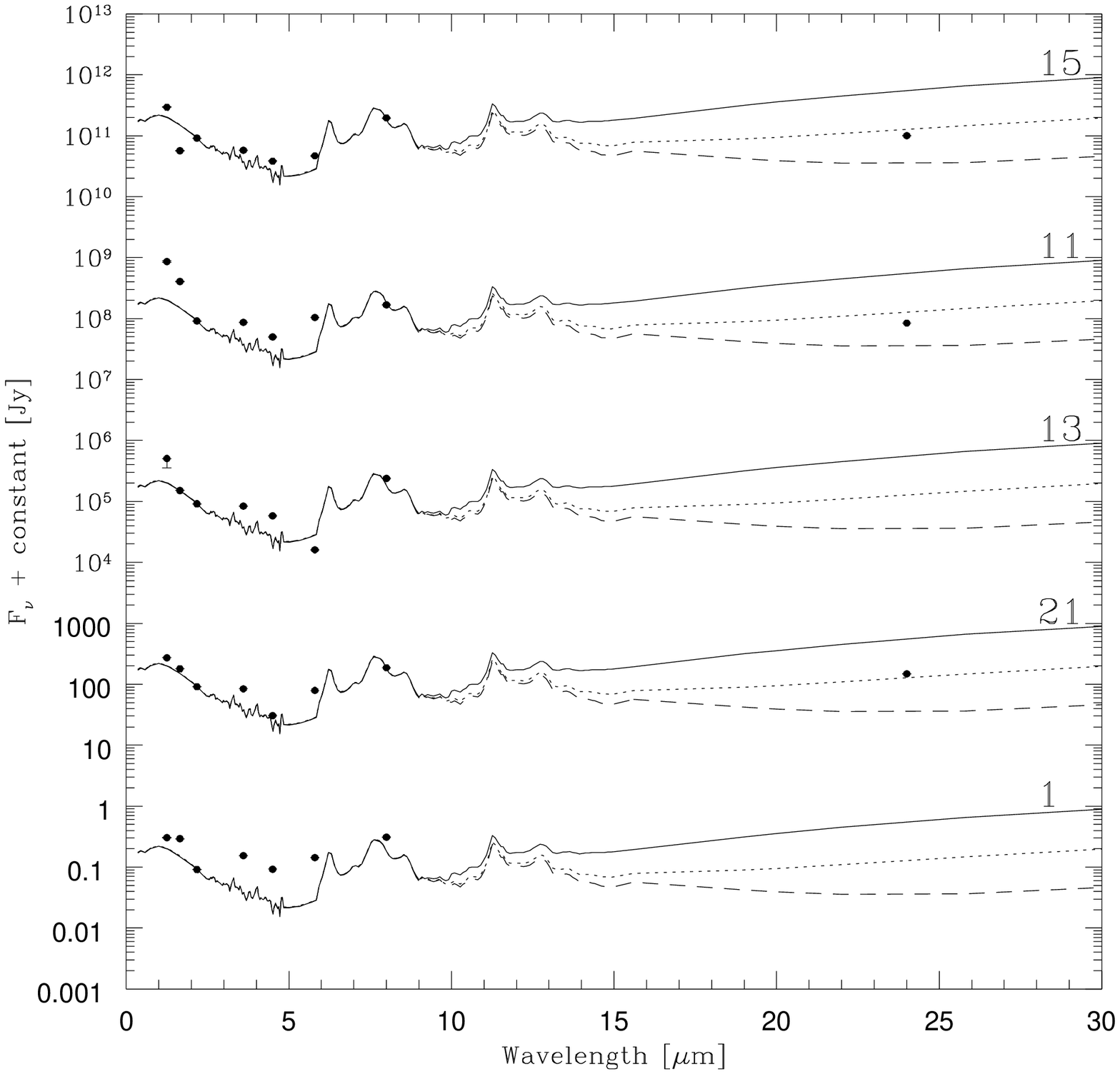}
}}
\caption{\label{fig:sed} The extinction-corrected SEDs, based on the
IRAS/DIRBE maps (Schlegel et al.\ 1998), of the twenty-five candidate
galaxies (see Table~\ref{tbl:phot}). The number associated with each
SED is the ID of the galaxy as listed in Table~\ref{tbl:coo}. From
{\em top} to {\em bottom}, the SEDs are ordered in decreasing Ks-band
magnitude and a constant is added to their measured flux densities to
allow for direct comparison. Models of Dale \& Helou (2002) with
$\alpha$~=~1.5 ({\em solid line}), 2.0 ({\em dotted line}), and 4.0
({\em dashed line}). The models have been normalized to the Ks
magnitude of each galaxy and shifted in the y-axis for easy
comparison.}
\end{figure}

\begin{figure}[h!]
\centerline{\hbox{
\includegraphics[width=130pt,height=130pt,angle=0]{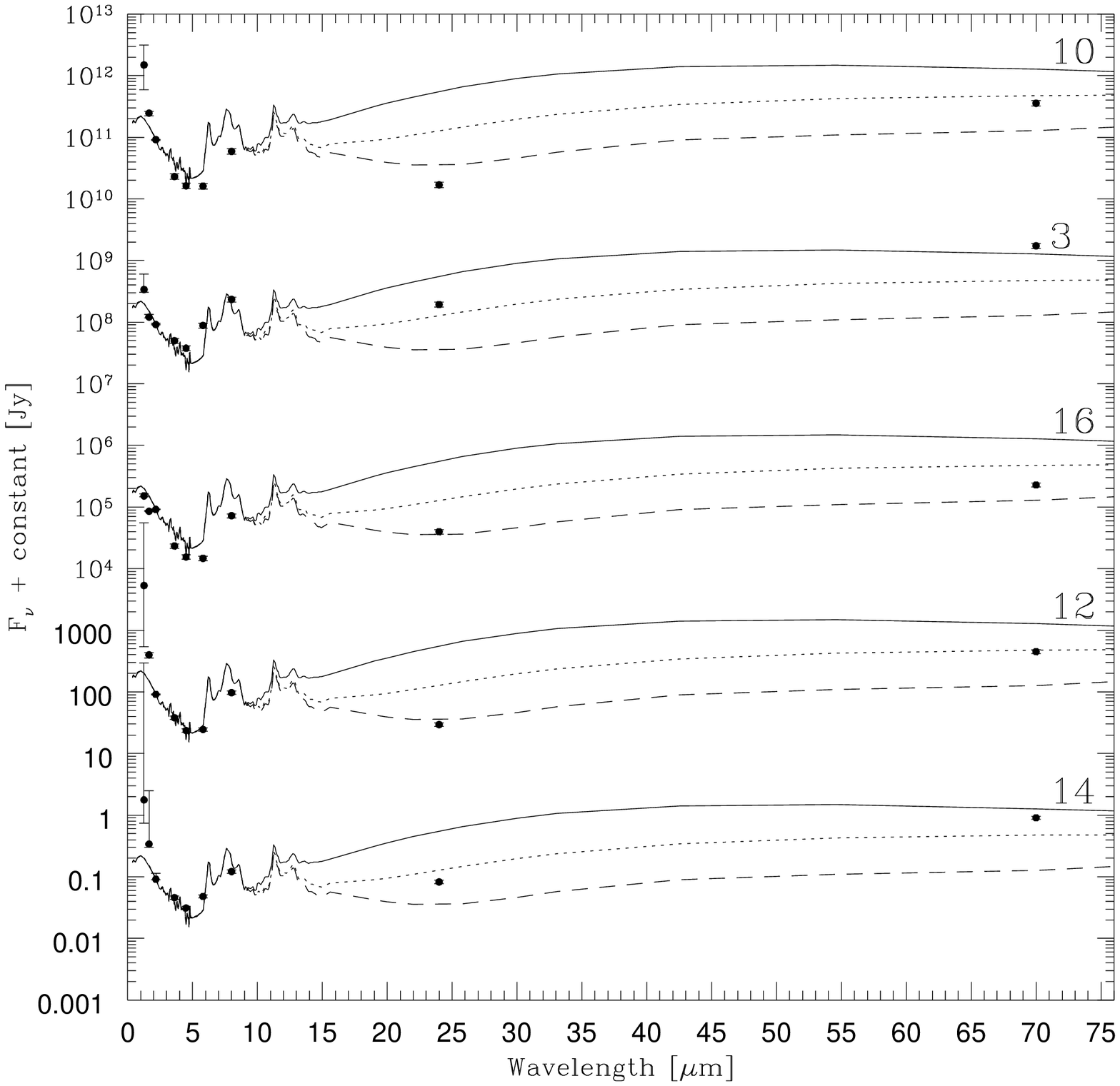}
\includegraphics[width=130pt,height=130pt,angle=0]{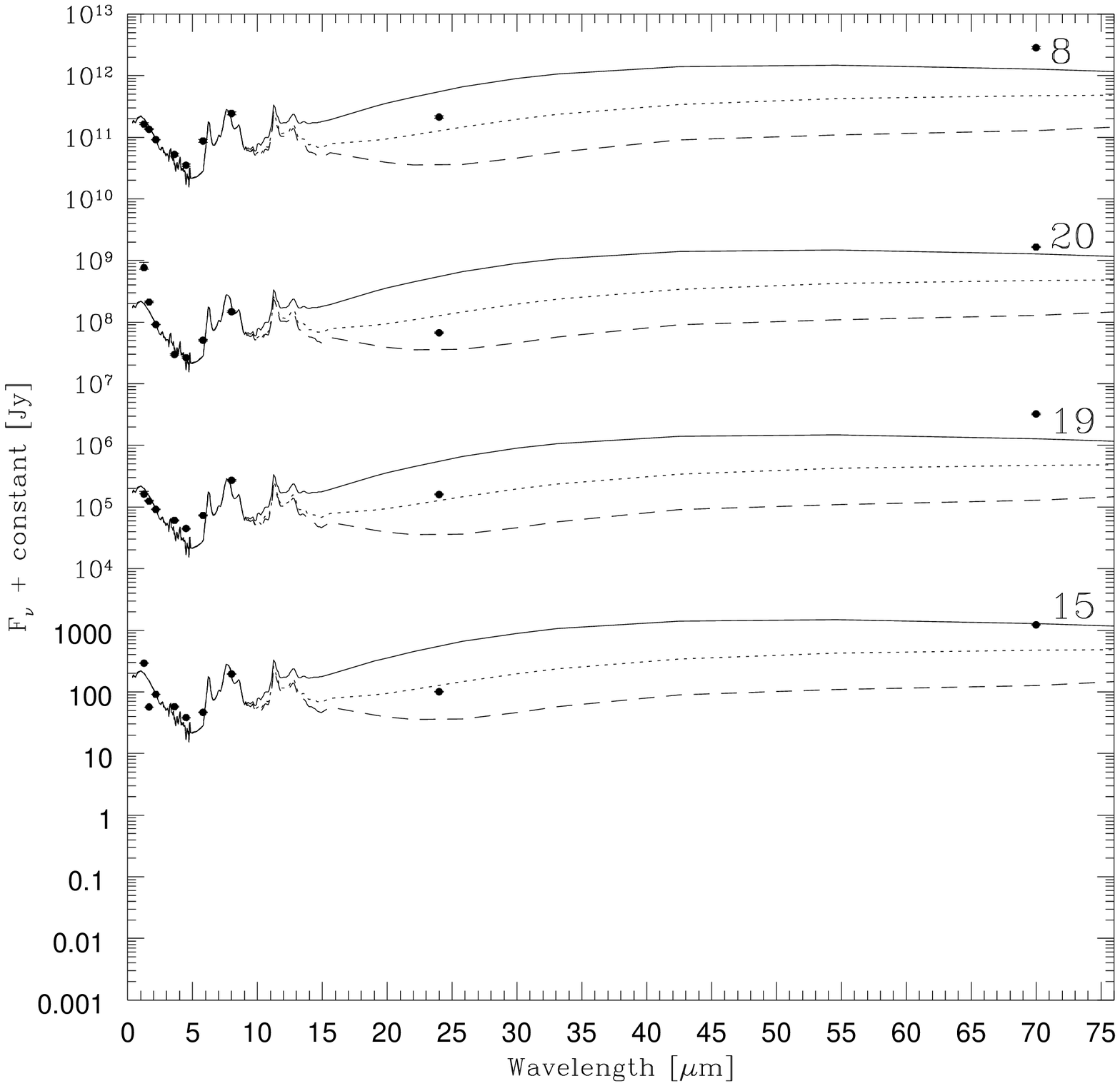}
}}
\caption{\label{fig:sed70} The extinction-corrected SEDs, based on the
IRAS/DIRBE maps (Schlegel et al.\ 1998), of galaxies with measured
70~\mum\ flux densities. The models shown are the same as in
\ref{fig:sed}.}
\end{figure}

The SEDs of the twenty-five candidate galaxies, covering the
wavelength range of 1.25 to 24~\mum, are shown in
Figure~\ref{fig:sed}. From top to bottom, the SEDs are ordered in
decreasing Ks-band magnitude and a constant is added to their measured
flux densities to allow for direct comparison. For nine of the
extra-galactic sources, a flux density was also measured in the
70~\mum\ images, and these extended SEDs are plotted in
Figure~\ref{fig:sed70}.

We selected the infrared SED distribution models of normal
star-forming galaxies of Dale et al. (2001) and Dale \& Helou (2002)
to create generic galaxy SED templates and compare them with the
observations. Our main reasons for selecting these models over other
ones were that they cover a wide wavelength range, from 3 to
1100~\mum\ and are described by a single parameter,
$f_{\nu}$(60~\mum)/$f_{\nu}$(100~\mum). The large wavelength
coverage, based on IRAS and ISO measurements (Dale \& Helou 2002), is
particularly important for this study as the new sources suffer from
significant extinction at shorter wavelengths, making the longer
wavelengths flux density measurements more reliable. The local SEDs in
these models are combined assuming a power-law distribution of dust
mass over heating intensity in a given galaxy, $dM_d(U) \propto
U^{-\alpha} dU$, where $dM_d(U)$ is the dust mass heated by a
radiation field at intensity $U$ and the exponent $\alpha$ is a
parameter that represents the relative contributions of the different
local SEDs. The range of $\alpha$ that describes the suite of normal
galaxy SEDs is approximately $1.0 < \alpha < 2.5$.  The models were
obtained from Dale's
website\footnote{http://faraday.uwyo.edu/$\sim$ddale/research/seds/seds.html}.
The file contained the model SEDs for a range of $\alpha$ of 0.0625 to
4.0 (step size of 0.0625).

Three model SEDs are shown in Figure~\ref{fig:sed} and \ref{fig:sed70}
along with the data points. These model SEDs were generated using a
value for the parameter $\alpha$ of 1.5, 2.0 and 4.0 and normalized to
the Ks flux density of each galaxy. They were chosen to approximately
mimic the range of behaviour of the measured SEDs. The first striking
result of this model-to-data comparison is that the SEDs of all our
candidate extra-galactic sources are unambiguously consistent with the
SEDs of normal galaxies. Moreover, we found that the majority of them,
i.e.\ twenty-three out of twenty-five, showed an extended disk at
8~\mum\ (see Figure~\ref{fig:8um}). This fraction drops to 50\% in the
near-infrared (see Figure~\ref{fig:2mass}) as the extinction becomes
larger at shorter wavelengths. The two sources that appeared compact
at 8~\mum, source ID 15 and 16 in Table~\ref{tbl:coo}, revealed their
disk-like morphology in their 2MASS images.

This selection bias was not a surprise as elliptical galaxies could
not be easily distinguished from compact Galactic sources in our
visual search given the spatial resolution of the IRAC and MIPS images
and their similar infrared colors. Assuming the morphological mix of
nearby galaxies observed in the field ($\sim$~25\% ellipticals; see,
e.g., van den Bergh 1998), we estimated that our search may have
missed $\sim$~8 early-type galaxies. Needless to say, this fraction
increases if the galaxies are in a cluster environment. However, given
the growing evidence that as much as $\sim$~70\% of the nearby
elliptical galaxies contain a significant amount of gas and dust (see,
e.g., Combes et al.\ 2007, Morganti et al.\ 2006, Jura 1986) in both
the field and cluster environments, our mid- to far-infrared search
probably only missed $\sim$~6 ($\sim$~18\%) dusty early-types.

\section{Identify Membership to Local Large-Scale Structure}

\subsection{Location in the 2-D Sky}

\begin{figure}[h!]
\centerline{\hbox{
\includegraphics[width=220pt,height=220pt,angle=0]{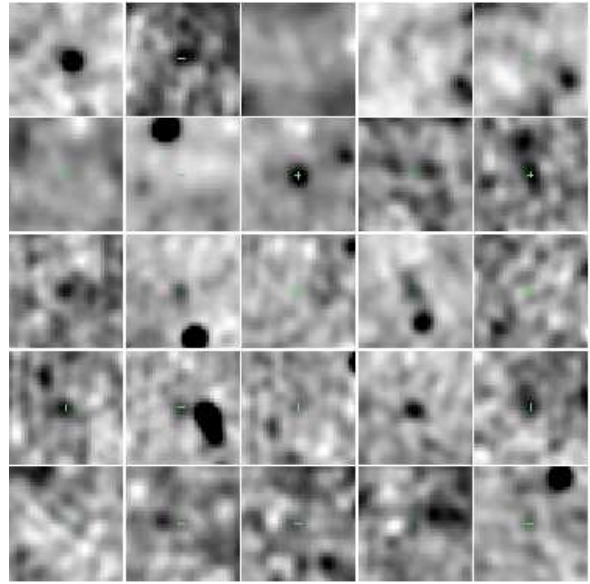}
}}
\caption{\label{fig:NVSS} NVSS 1.4~GHz postage stamps of the same
galaxies. The galaxies are displayed in the same order as shown in
Figure~\ref{fig:8um}. The FOV of each image is 5.5\arcmin\ $\times$
5.5\arcmin\ (7.5\arcsec\ per pixel) with N up and E to the left. The
{\em green} crosses overlayed on the images represent the central
location of the candidates as listed in Table~\ref{tbl:coo}. The
images are displayed in reserve colormap and the scale parameters of
the postage stamps were optimized for each individual source and
therefore differ for each image.}
\end{figure}

In order to ascertain the location of the newly discovered galaxies
within the two-dimensional local large-scale structure, we first
queried the NASA Extragalactic Database (NED) to cross-check these new
sources within the existing database and to also identify any
previously detected galaxies listed for our survey region ($l = 40 -
65$\arcdeg, $|b| \lesssim 1$\arcdeg\ and $z < 0.08$). The outcome of
this search was negative, and no matches were found. However, five
additional sources, with no match to ours, were returned by NED and
identified as detections from the HI Parkes Zone of Avoidance Survey
(HIZOA; Donley et al.\ 2005) and the HI Parkes All-Sky Survey (HIPASS;
Staveley-Smith et al.\ 2000). Taking into account the 3$\sigma$
positional uncertainty of the HI Parkes surveys of 3\arcmin, we
searched for the near- to far-infrared counterparts to these sources
and found no extended sources at these locations. As this first query
returned only sources with redshift, we queried NED again at each of
the candidate galaxy location without any redshift constraint and
found this time seven matches (see Table~\ref{tbl:coo}), four
classified as detections from the NRAO/VLA Sky Survey (NVSS), two from
the Westerbork Synthesis Radio Telescope Galactic Plane Survey
(WSRTGP) and one from the First White+Giommi+Angelini ROSAT X-Ray
sources list (1WGA). The NVSS postage stamps of all the candidates are
shown in Figure~\ref{fig:NVSS}. We carefully examined all NVSS postage
stamps and confirmed the four NVSS and two WSRTGP detections returned
by NED as well as an additional five detections, bringing the total to
eleven sources with believeable detections at 1.4~GHz. The 1WGA source
returned from NED did not have a detection at 1.4~GHz.

\begin{figure}[h!]
\centerline{\hbox{
\includegraphics[width=220pt,height=220pt,angle=0]{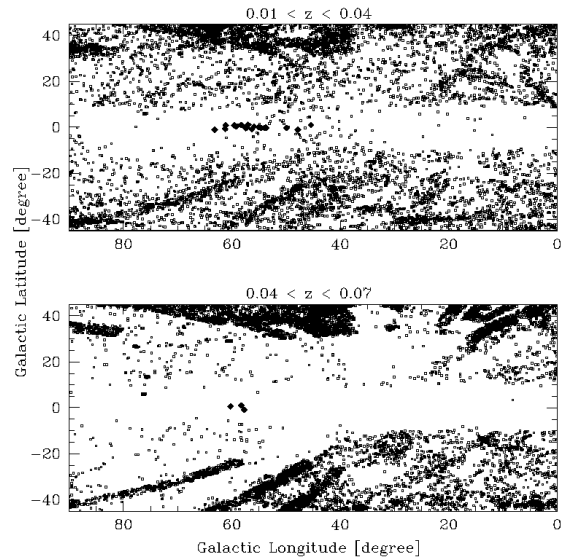}
}}
\caption{\label{fig:LSS} Spatial distribution of the new galaxies
({\em filled lozenges} near the Galactic equator) in the 40\arcdeg
$\ge$ l $\ge$ 65\arcdeg\ region superimposed on the distribution of
NED galaxies ({\em open squares}) in the redshift range $0.01 < z <
0.04$ ({\em top} panel) and $0.04 < z < 0.07$ ({\em bottom}
panel). The figure, more specifically the {\em top panel}, suggests
that the new galaxies are consistent with being spatially located on
an extension of the already known large-scale structure at higher
Galactic latitude.}
\end{figure}

We then looked at the location of the discovered galaxies within the
two-dimensional distribution of the known galaxies in the local
Universe, i.e.\ we extended our search to a larger area in the sky ($l
= 0 - 90$\arcdeg, $|b| \leq 45$\arcdeg) and within a redshift space of
$z = 0.01 - 0.04$ (11243 galaxies) and $z = 0.04 - 0.07$ (13055
galaxies). These included the HIZOA galaxies found by Donley et al.\
(2005). The position of the new galaxies are overlaid on the map of
the local large-scale structure in Figure~\ref{fig:LSS}, top and
bottom panel, respectively. We find that these overdensities of
galaxies are located in the sky near a local large-scale structure,
seen also in the 2MASS extended source catalog (Jarrett 2004) for
redshifts $z = 0.01 - 0.03$, and extending from $|b| = 4 -
40$\arcdeg. Our findings therefore provide the first strong evidence
of a bridging at very low Galactic latitude between the two
large-scale structures on both sides of the Galactic plane.

\subsection{Distance/Redshift Determination}

Although the overdensities of galaxies which we have discovered appear
to be near a filament, this does not necessarily prove their
membership.  However, a photometric redshift estimate can certainly
narrow down the probability that these discovered galaxies are part of
this local super-structure.

We were able to estimate the distance and therefore redshift to our
candidate galaxies using two independent methods. The first method
simply made use of the Ks-band magnitudes of the sources and assumed
the spectral energy distribution of an L$^{\ast}$ galaxy, with
M$_K^{\ast} = -24$~mag obtained from the K-band field luminosity
function (Kochanek et al.\ 2001). The SEDs normalized to the Ks-band
are shown in Figure~\ref{fig:sed}. The inferred distances are 54 to
356~Mpc. This puts our galaxies in the $4050 - 26700$~km/s velocity
range, or at a redshift of $0.013 - 0.084$ (in our calculations, we
assumed a cosmology with $H_0 = 75$~km~s$^{-1}$~Mpc$^{-1}$, $\Omega_M
= 0.3$, and $\Omega_{\Lambda} = 0.7$).

\begin{figure}[h!]
\centerline{\hbox{
\includegraphics[width=130pt,height=130pt,angle=0]{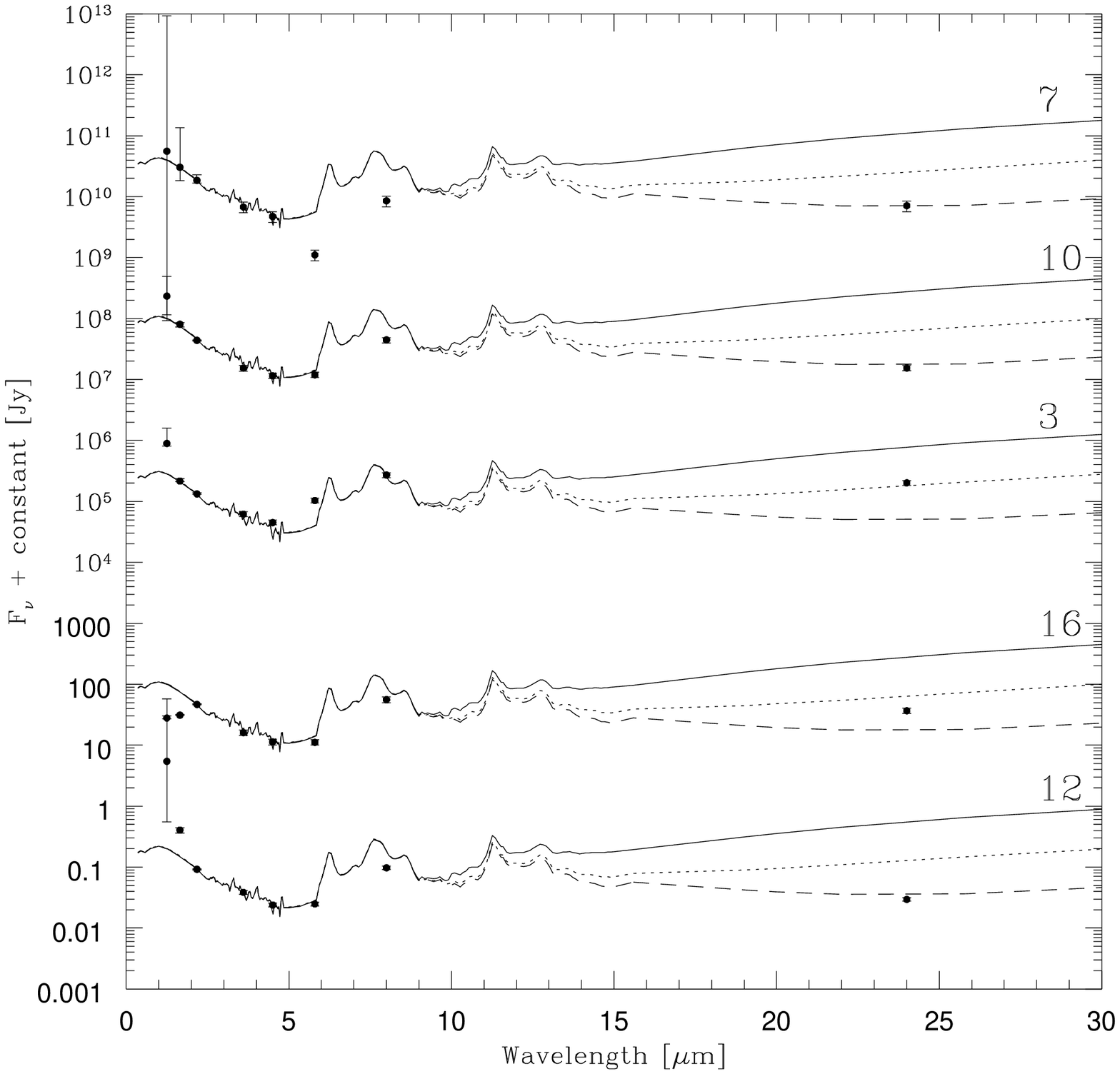}
\includegraphics[width=130pt,height=130pt,angle=0]{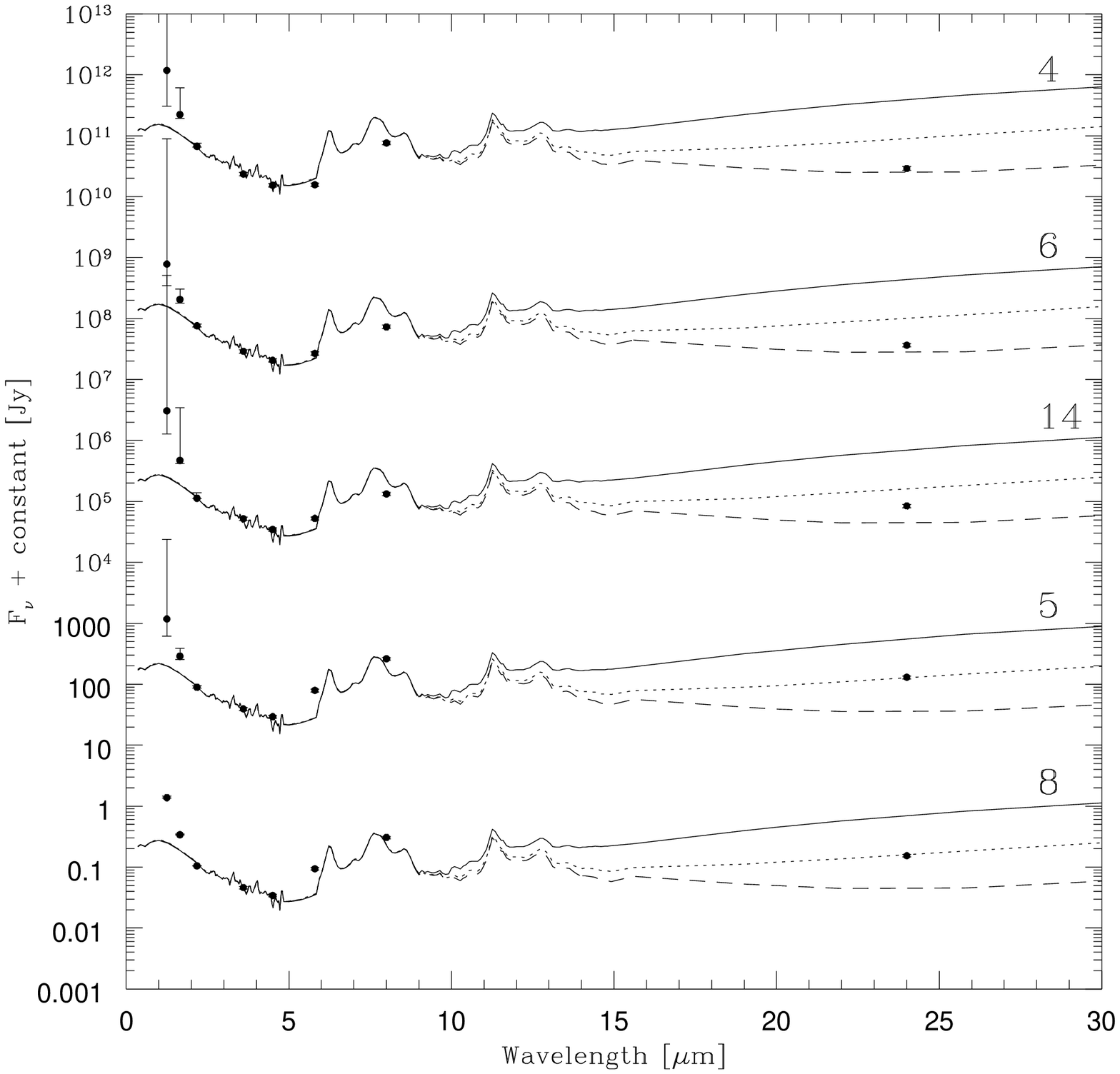}
}}
\centerline{\hbox{
\includegraphics[width=130pt,height=130pt,angle=0]{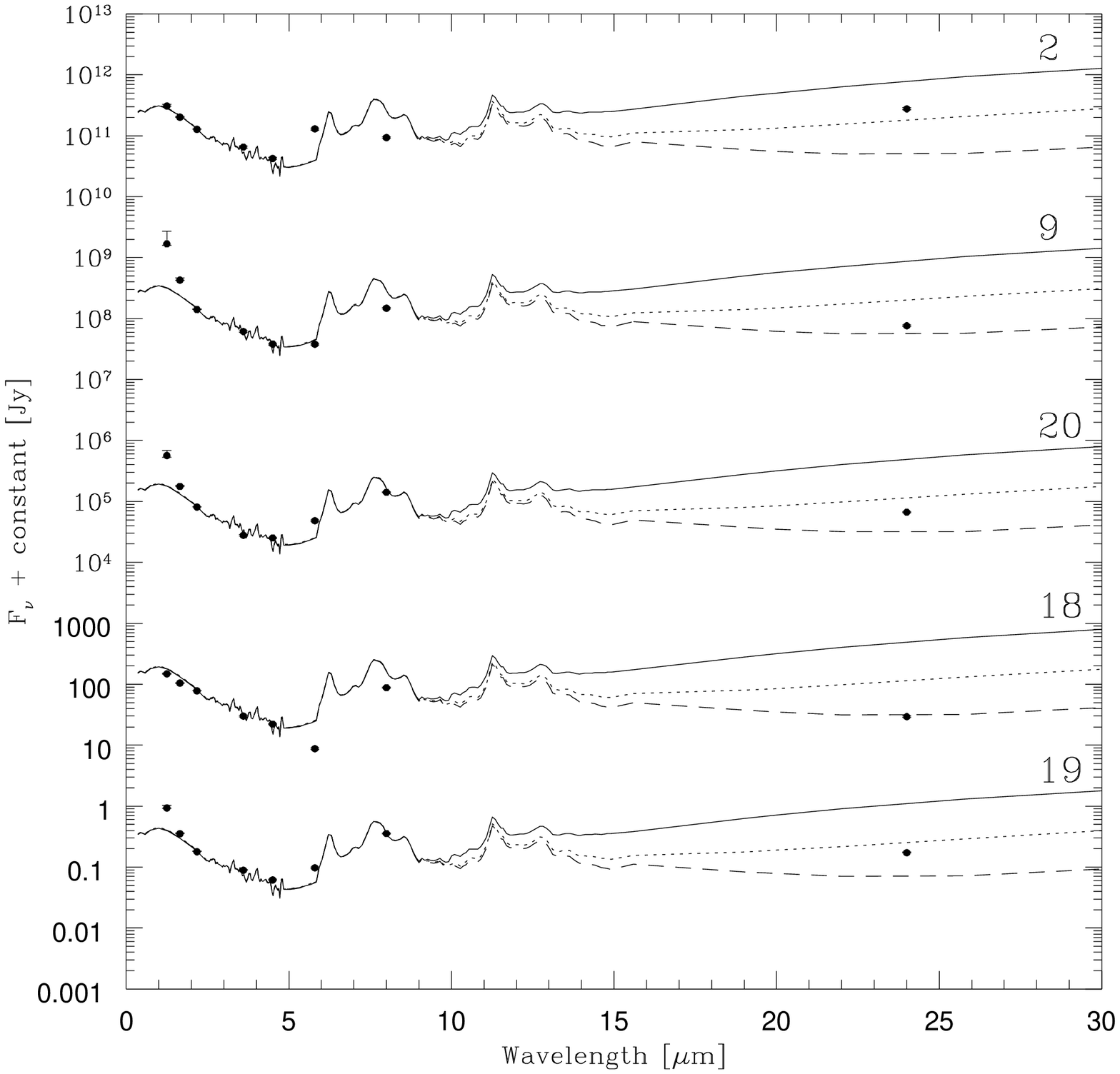}
\includegraphics[width=130pt,height=130pt,angle=0]{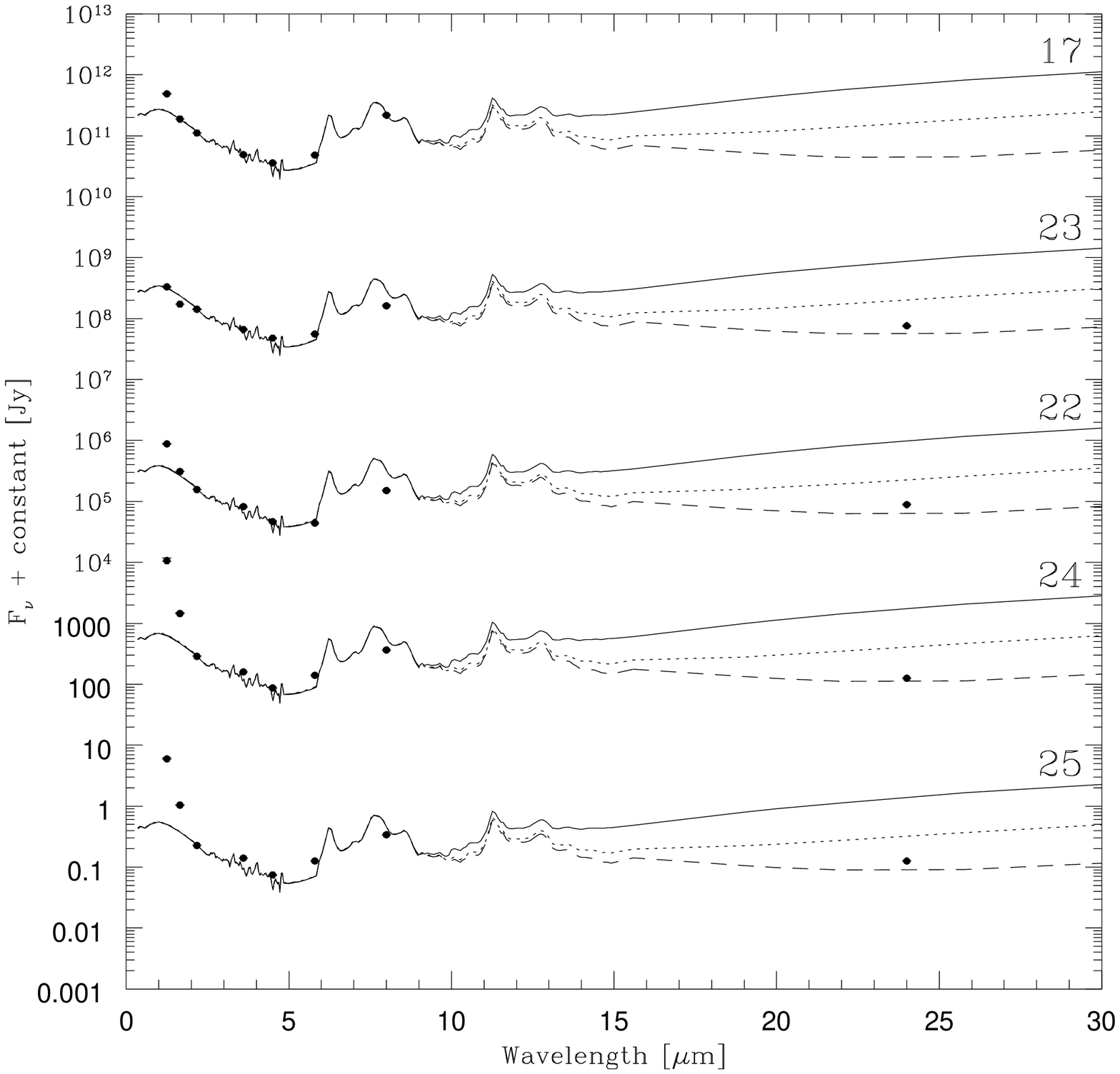}
}}
\centerline{\hbox{
\includegraphics[width=130pt,height=130pt,angle=0]{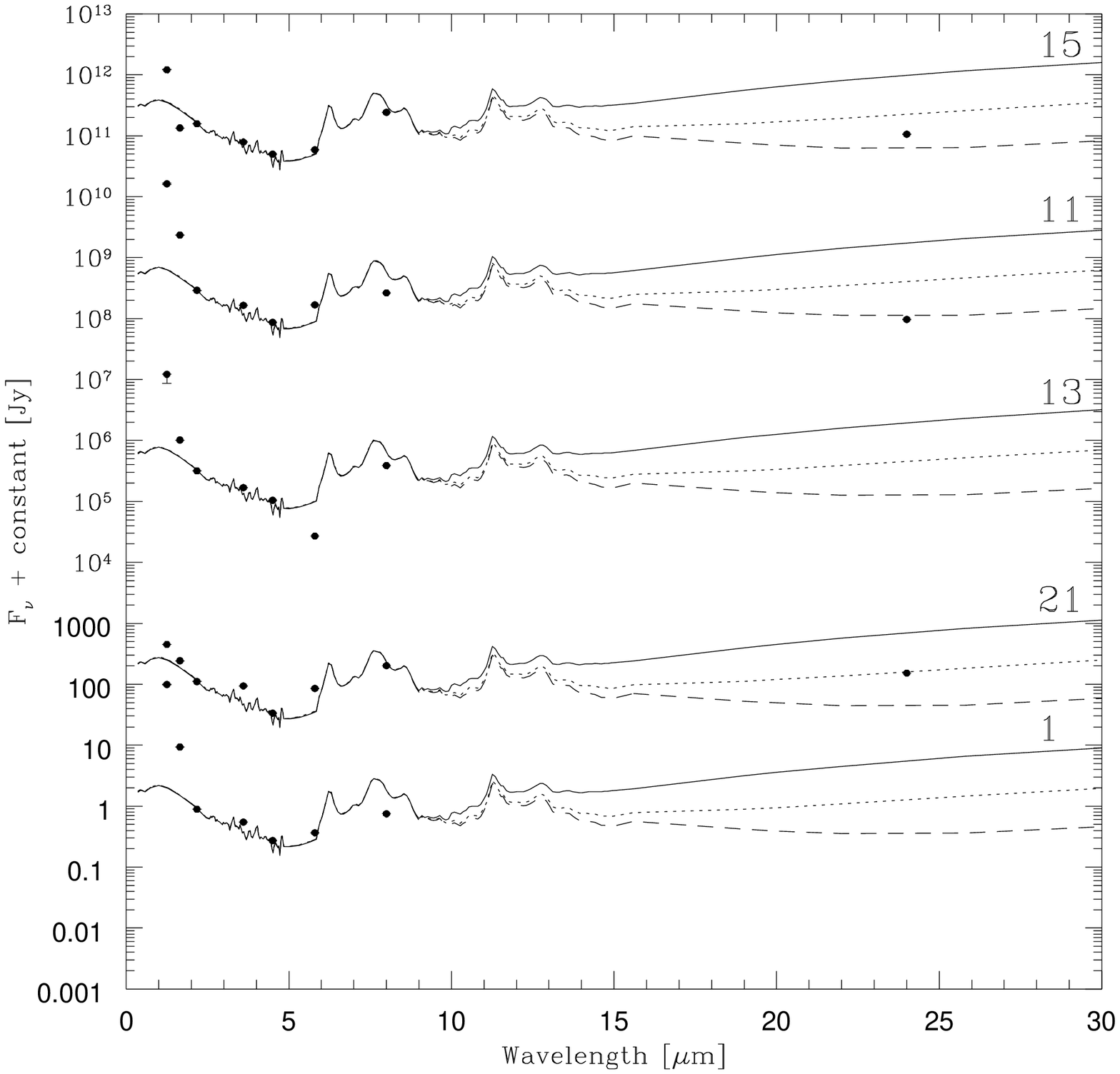}
}}
\caption{\label{fig:sedIRAC} The extinction-corrected SEDs, based on
our SED fit, of the twenty-five candidate galaxies (see
Table~\ref{tbl:photsed}). The models shown are the same as in
Figure~\ref{fig:sed}. The models have been normalized to the IRAC 3.6
and 4.5~\mum\ bands.}
\end{figure}

The second and more robust method made use of the full SED, weighted
most heavily on the IRAC channel 1 and 2 flux density measurements,
which suffered the least from uncertainties associated with extinction
(more important at shorter wavelengths) and PAHs emission features
(more important in IRAC channel 3 and 4), to estimate the distance to
the galaxies. The new fits, weighted to the IRAC channel 1 and 2, are
shown in Figure~\ref{fig:sedIRAC}, along with the extinction-corrected
flux densities obtained using the SED fit (see
Table~\ref{tbl:photsed}). The redshifts and new values for the visual
extinction derived using this more robust method are listed in
Table~\ref{tbl:z} and Table~\ref{tbl:ext} (last column),
respectively. The galaxies found in this region of the sky have
estimated redshifts ranging between 0.016 and 0.068 (distance of 66 to
288~Mpc).

As seen in Table~\ref{tbl:ext} (first and last column) the A(V) values
derived from the SED fits are sometimes significantly different than
the IRAS/DIRBE extinction values. For eight of the twenty-five galaxy
candidates (source ID 4, 5, 6, 7, 10, 16, 18, and 20), the A(V) values
derived from the SED fits are lower than the ones given by Schlegel et
al.\ (1998) by a median value of 2.8~mag, with the largest difference
being the value of 14.7~mag (source ID 7). A similar IRAS/DIRBE
overestimate of the Galactic visual extinction values was reported by
Nagayama et al.\ (2004) and Schr\"oder et al.\ (2007) in the direction
of the highly obscured radio-bright galaxy PKS~1343-601 at Galactic
coordinates of ($l$,$b$)~=~(309.7\arcdeg,1.7\arcdeg). However, our
comparison indicates that for the remaining seventeen galaxies, the
A(V) value derived is larger than the IRAS/DIRBE value, by a median
value of 4.0~mag and with the largest difference being the value of
20.8~mag (source ID 1). This suggests that the IRAS/DIRBE maps may be
underestimating the amount of extinction for the most obscured regions
($|b| \leq 1$\arcdeg) of the Milky Way.

As can be seen by looking at the postage stamps, our candidate
galaxies are all fainter (after the flux densities have been corrected
for dust extinction) than the two galaxies discussed in Jarrett et
al.\ (2007). Indeed, the IRAC channel 1 flux densities of our sample
range from 2.4 to 22.8~mJy (see Table~\ref{tbl:phot}) whereas the two
galaxies discussed in Jarrett et al. (2007) have flux densities of
35.0 and 18.3~mJy (see their Table 1). Therefore, our galaxies probe
flux densities ten times fainter than Jarrett et al.\ (2007). It
follows that we are probing deeper in redshift space, $z = 0.016 -
0.068$ (see Table~\ref{tbl:z}), beyond the $z \simeq 0.015$ galaxies
of Jarrett et al.\ (2007). It is therefore not surprising that for the
majority of the candidate galaxies, we did not find an HI
counterpart. By comparing the properties of our galaxy with largest
IRAC channel 1 and 2 flux densities, source ID 3 (22.8 and 17.2~mJy,
see Table~\ref{tbl:phot}), with Jarrett et al.'s two galaxies (G1:
35.0 and 24.1~mJy, G2: 18.3 and 12.7~mJy, see their Table 1), we find
that our photometric redshift estimate of $z = 0.016$ for this galaxy
is entirely consistent with the HI radial velocity redshifts derived
for Jarrett et al.'s galaxies ($z \simeq 0.015$).

\begin{figure}[h!]
\centerline{\hbox{
\includegraphics[width=220pt,height=220pt,angle=0]{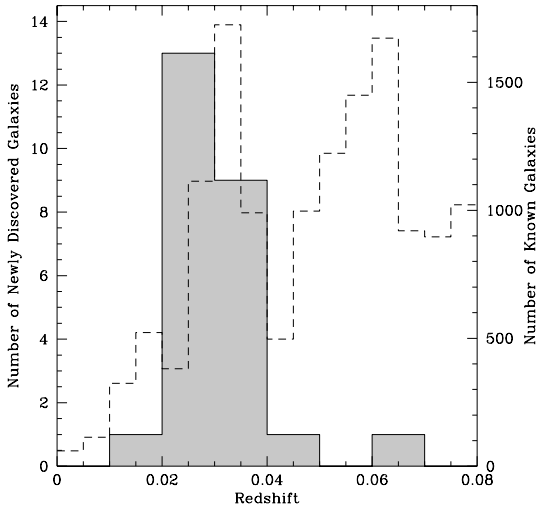}
}}
\caption{\label{fig:zhist} {\em Left axis}: Redshift histogram ({\em
grey shaded} and {\em solid line}) of our twenty-five newly discovered
galaxies. {\em Right axis}: Redshift histogram ({\em dashed line}) of
all known galaxies from NED in the region of the Galactic plane with
$40\arcdeg < l < 70\arcdeg$ and $-45\arcdeg < b < 45\arcdeg$ with $0 <
z < 0.08$. The figure suggests that the new galaxies are consistent in
redshift space with an extension of the already known large-scale
structure at higher Galactic latitude.}
\end{figure}

\begin{figure}[h!]
\centerline{\hbox{
\includegraphics[width=220pt,height=160pt,angle=0]{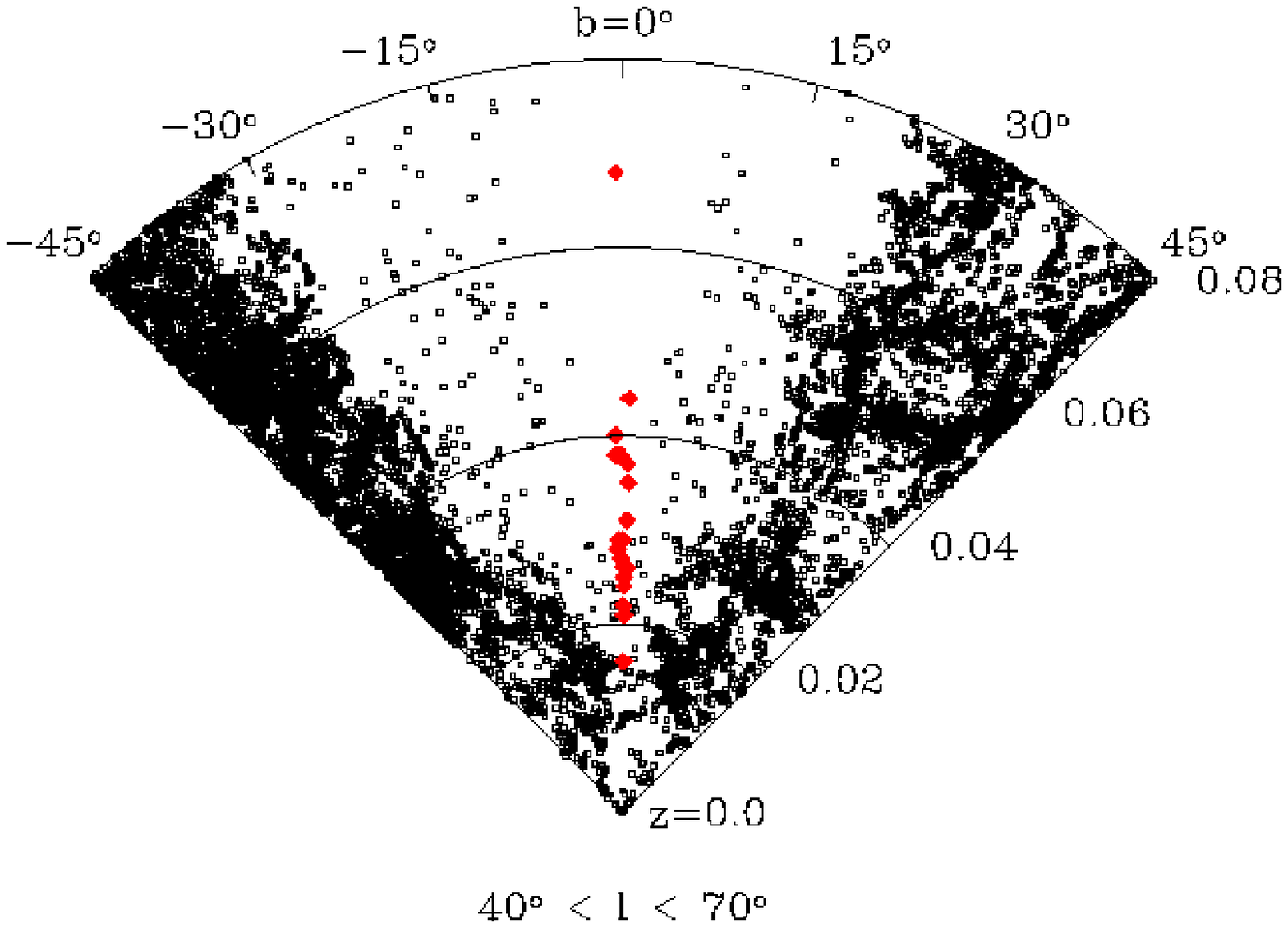}
}}
\centerline{\hbox{
\includegraphics[width=220pt,height=160pt,angle=0]{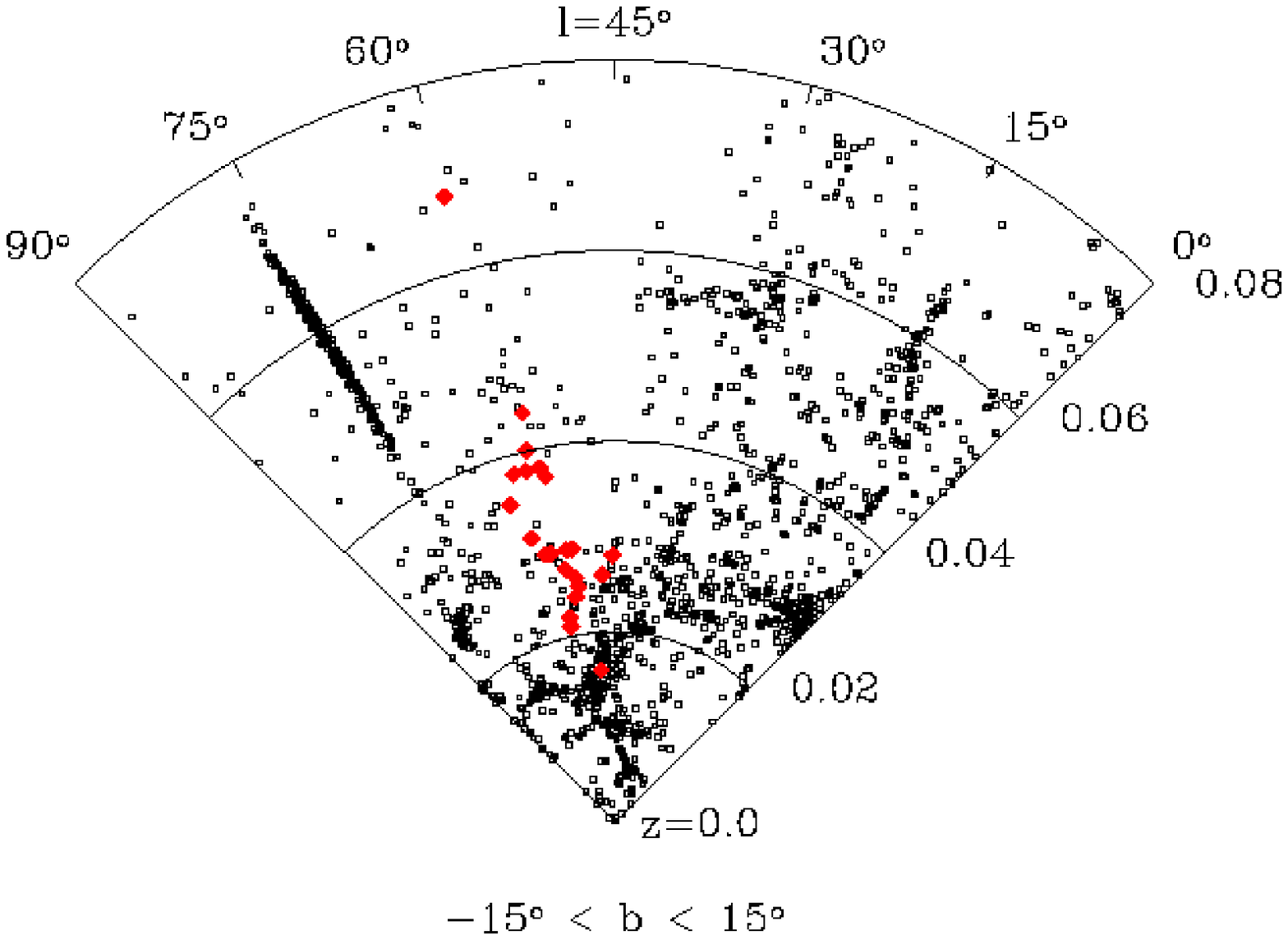}
}}
\caption{\label{fig:zpie} A cone plot in redshift space of the NED
galaxies ({\em open squares}) located in a slice of $-45\arcdeg < b <
45\arcdeg$ for Galactic longitudes in the range $40\arcdeg < l <
70\arcdeg$ ({\em top} panel), including the new galaxies of this study
({\em filled red lozenges}). The {\em bottom} panel shows the cone diagram
with $0\arcdeg < l < 90\arcdeg$ for Galactic latitudes in the range
$-15\arcdeg < b < 15\arcdeg$. The overall distribution of these new
objects is consistent with the local super-structures found between
the redshifts of $z = 0 - 0.08$.}
\end{figure}

A histogram of the redshifts of the twenty-five discovered galaxies is
shown in Figure~\ref{fig:zhist}. We find that the majority of them
occupy the redshift range $z \simeq 0.01 - 0.05$, with one source
located at $z \simeq 0.07$. For comparison, the redshift distribution
of all known galaxies in the redshift range $z = 0.0 - 0.08$ and with
$l = 40 - 70$\arcdeg\ and $|b| \leq 45$\arcdeg, i.e.\ 13907 galaxies,
is also shown in Figure~\ref{fig:zhist}. The overdensity in redshift
space of our sample agrees remarkably well with the one of the
previously known local large-scale structure in this region (see
Figure~\ref{fig:zpie}). This seems to agree again with the
interpretation that these newly revealed galaxies belong to an
extension to lower Galactic latitude of the already known local
super-structure and belong to a bridge extending the filamentary
structure seen on each side of the Galactic plane. However, remember
that our redshift determination depends on the assumption that these
are all L$^{\ast}$ galaxies. If they are not, then clearly their
redshift distribution will change. For example, it is possible that
the galaxy with the largest redshift is not an L$^{\ast}$ galaxy but
somewhat fainter and that we are overestimating its distance.

\section{Summary}

We reported in this paper the discovery of twenty-five galaxies in
the ZoA located around $l \sim 47$ and 55\arcdeg\ and $|b| \lesssim
1$\arcdeg\ in the Sagitta-Aquila region. These overdensities are
consistent with the local large-scale structure found at similar
Galactic longitude and extending from $|b| \sim 4$ to 40\arcdeg. We
presented their SEDs and provided evidence of their extra-galactic
nature. Their redshifts were estimated using the infrared photometric
measurements obtained from 2MASS and Spitzer. We found that the
majority of the newly discovered galaxies occupy the redshift range 
$z \simeq 0.01 - 0.05$, with one source located at $z \simeq 0.07$. 
Comparison with known sources in the local Universe revealed that 
these galaxies are located at similar overdensities in redshift space. 
This seems to imply that these newly discovered galaxies belong to 
an extension to lower Galactic latitude of the already known local 
super-structure, providing a first view of the bridging between both 
sides of the Galactic plane.

However, we must emphasize that these photometric redshifts, which
suffer from the uncertainties associated with the extinction
correction and are based on the assumption that these are all
L$^{\ast}$ galaxies, can only be determined with certainty using
spectroscopic data. Therefore, we are carrying out follow-up
near-infrared spectroscopic observations of the newly discovered
galaxies with ground-based telescopes. The spectroscopic redshifts
will allow us to derive a better estimate of the amount of extinction 
in the direction of our sources, which is highly uncertain. 
In addition, with their distance fixed, we will be able to calculate 
their intrinsic luminosities and estimate their stellar masses.

In conclusion, the work presented here clearly demonstrates the power
of mid- to far-infrared surveys in finding galaxies in highly obscured
regions such as the plane of our Galaxy, surveying even the lowest
Galactic latitudes of $|b| \lesssim 1$\arcdeg. Moreover, the relative
success of our survey implies that the ZoA should no longer be avoided
and further studies should be undertaken to expand our knowledge of
the local large-scale structure in this heavily obscured part of the
Universe.

\acknowledgements

We thank an anonymous referee for helpful suggestions on improving the
manuscript. This work is based in part on observations made with the
{\em Spitzer Space Telescope}, which is operated by the Jet Propulsion
Laboratory, California Institute of Technology, under NASA contract
1407.

\newpage

\begin{deluxetable}{rlrrrr}
\tablecaption{Coordinates of Spitzer Galaxies in the ZoA \label{tbl:coo}}
\tablewidth{0pt}
\tablecolumns{6}
\tablehead{ \colhead{ID}  &\colhead{Name}  &\colhead{$\alpha_{J2000}$}  &\colhead{$\delta_{J2000}$}  &\colhead{ \gall } &\colhead{ \galb } }
\startdata
1   &SPITZER191050+113409\tablenotemark{a}  &19h10m50.432s   &11d34m09.38s  &45.42353  & 1.01550\\
2   &SPITZER192252+124513  &19h22m52.169s   &12d45m12.84s  &47.84276  &-1.03199\\
3   &SPITZER192404+145632  &19h24m04.473s   &14d56m31.75s  &49.91004  &-0.25584\\
4   &SPITZER193206+183140  &19h32m06.060s   &18d31m39.88s  &53.97787  &-0.23159\\
5   &SPITZER193145+181940\tablenotemark{b}  &19h31m44.840s   &18d19m39.88s  &53.76221  &-0.25451\\
6   &SPITZER193133+181516  &19h31m32.750s   &18d15m16.45s  &53.67504  &-0.24784\\
7   &SPITZER193105+182614  &19h31m04.748s   &18d26m14.12s  &53.78187  &-0.06284\\
8   &SPITZER193337+190727\tablenotemark{c}  &19h33m36.564s   &19d07m27.23s  &54.67258  &-0.25585\\
9   &SPITZER193320+193334  &19h33m20.301s   &19d33m34.38s  &55.02256  & 0.01080\\
10  &SPITZER193349+202958\tablenotemark{d}  &19h33m48.593s   &20d29m58.49s  &55.89871  & 0.36900\\
11  &SPITZER193915+201426  &19h39m14.703s   &20d14m25.88s  &56.29427  &-0.86925\\
12  &SPITZER193825+212701  &19h38m24.972s   &21d27m01.29s  &57.25305  &-0.10634\\
13  &SPITZER193324+214918  &19h33m24.468s   &21d49m17.93s  &57.01030  & 1.09097\\
14  &SPITZER193956+213017\tablenotemark{e}  &19h39m56.106s   &21d30m17.31s  &57.47421  &-0.38725\\
15  &SPITZER194223+212514  &19h42m23.331s   &21d25m13.72s  &57.68324  &-0.92517\\
16  &SPITZER193633+225125  &19h36m32.781s   &22d51m25.18s  &58.26802  & 0.95949\\
17  &SPITZER193608+225054\tablenotemark{f}  &19h36m07.572s   &22d50m54.33s  &58.21334  & 1.03982\\
18  &SPITZER193632+225355  &19h36m31.917s   &22d53m54.81s  &58.30269  & 0.98266\\
19  &SPITZER193958+235251\tablenotemark{g}  &19h39m58.334s   &23d52m50.69s  &59.54589  & 0.77617\\
20  &SPITZER194042+232429  &19h40m42.470s   &23d24m29.33s  &59.21788  & 0.39649\\
21  &SPITZER194211+242507  &19h42m10.975s   &24d25m06.66s  &60.26274  & 0.60358\\
22  &SPITZER194257+251849  &19h42m57.449s   &25d18m49.18s  &61.12754  & 0.89557\\
23  &SPITZER194349+251449  &19h43m48.966s   &25d14m49.40s  &61.16589  & 0.69407\\
24  &SPITZER194919+243738  &19h49m19.339s   &24d37m37.64s  &61.25773  &-0.69658\\
25  &SPITZER195443+260745  &19h54m43.154s   &26d07m45.08s  &63.17221  &-0.97524\\
\enddata
\tablenotetext{a}{NVSS J191049+113403}
\tablenotetext{b}{1WGA J1931.7+1818}
\tablenotetext{c}{NVSS J193336+190723}
\tablenotetext{d}{NVSS J193349+203131}
\tablenotetext{e}{WSRTGP 1937+2122}
\tablenotetext{f}{WSRTGP 1933+2243}
\tablenotetext{g}{NVSS J193958+235243}
\end{deluxetable}

\begin{deluxetable}{rrrrrrrrrrrc}
\tablecaption{Extinction Correction for Photometric Measurements of Spitzer Galaxies in the ZoA \label{tbl:ext}}
\tablewidth{0pt}
\tablecolumns{12}
\tablehead{ \colhead{ID} &\colhead{A(V)\tablenotemark{a}} &\colhead{A(J)} &\colhead{A(H)} &\colhead{A(Ks)} &\colhead{A(I1)} &\colhead{A(I2)} &\colhead{A(I3)} &\colhead{A(I4)} &\colhead{A(M1)} &\colhead{A(M2)} &\colhead{A(V) (SED fit)}\\
      &\colhead{0.5\mum} &\colhead{1.25\mum} &\colhead{1.65\mum} &\colhead{2.17\mum} &\colhead{3.6\mum} &\colhead{4.5\mum} &\colhead{5.8\mum} &\colhead{8.0\mum} &\colhead{24.0\mum} &\colhead{70.0\mum} &\colhead{0.5\mum} }
\startdata
1   &  8.9  & 2.7  & 1.6  & 1.0  & 0.6  & 0.5  & 0.4  & 0.4  & 0.1  & 0.02  &29.8 \\
2   & 14.2  & 4.3  & 2.6  & 1.7  & 0.9  & 0.8  & 0.7  & 0.7  & 0.2  & 0.03  &17.3 \\
3   & 22.6  & 6.8  & 4.1  & 2.7  & 1.5  & 1.3  & 1.1  & 1.1  & 0.3  & 0.04  &26.1 \\
4   & 23.0  & 6.9  & 4.1  & 2.7  & 1.5  & 1.3  & 1.1  & 1.1  & 0.3  & 0.04  &20.2 \\
5   & 17.1  & 5.2  & 3.1  & 2.0  & 1.1  & 1.0  & 0.8  & 0.8  & 0.2  & 0.03  &16.9 \\
6   & 19.5  & 5.9  & 3.5  & 2.3  & 1.3  & 1.1  & 1.0  & 0.9  & 0.3  & 0.04  &17.8 \\
7   & 31.0  & 9.3  & 5.6  & 3.7  & 2.0  & 1.7  & 1.5  & 1.5  & 0.4  & 0.06  &16.3 \\
8   & 13.8  & 4.2  & 2.5  & 1.6  & 0.9  & 0.7  & 0.7  & 0.7  & 0.2  & 0.02  &15.9 \\
9   & 17.6  & 5.3  & 3.2  & 2.1  & 1.2  & 1.0  & 0.8  & 0.8  & 0.2  & 0.03  &21.6 \\
10  & 23.8  & 7.2  & 4.3  & 2.8  & 1.6  & 1.3  & 1.2  & 1.1  & 0.3  & 0.04  &17.1 \\
11  & 11.6  & 3.5  & 2.1  & 1.4  & 0.8  & 0.7  & 0.6  & 0.5  & 0.2  & 0.02  &22.2 \\
12  & 28.2  & 8.5  & 5.1  & 3.3  & 1.9  & 1.6  & 1.4  & 1.3  & 0.4  & 0.05  &28.3 \\
13  & 14.5  & 4.3  & 2.6  & 1.7  & 1.0  & 0.8  & 0.7  & 0.7  & 0.2  & 0.03  &26.0 \\
14  & 21.7  & 6.6  & 3.9  & 2.6  & 1.4  & 1.2  & 1.1  & 1.0  & 0.3  & 0.04  &23.7 \\
15  & 11.9  & 3.6  & 2.2  & 1.4  & 0.8  & 0.7  & 0.6  & 0.5  & 0.2  & 0.02  &17.0 \\
16  & 11.6  & 3.5  & 2.1  & 1.4  & 0.8  & 0.6  & 0.6  & 0.5  & 0.2  & 0.02  &5.5  \\
17  & 10.8  & 3.2  & 1.9  & 1.3  & 0.7  & 0.6  & 0.5  & 0.5  & 0.1  & 0.02  &12.6 \\
18  & 10.7  & 3.2  & 1.9  & 1.3  & 0.7  & 0.6  & 0.5  & 0.5  & 0.1  & 0.02  &9.3  \\
19  &  9.6  & 2.9  & 1.7  & 1.1  & 0.6  & 0.5  & 0.5  & 0.4  & 0.1  & 0.02  &15.9 \\
20  & 15.7  & 4.7  & 2.8  & 1.8  & 1.0  & 0.9  & 0.8  & 0.7  & 0.2  & 0.03  &14.6 \\
21  & 11.4  & 3.4  & 2.0  & 1.3  & 0.8  & 0.6  & 0.6  & 0.5  & 0.2  & 0.02  &13.2 \\
22  &  7.8  & 2.3  & 1.4  & 0.9  & 0.5  & 0.4  & 0.4  & 0.4  & 0.1  & 0.01  &12.9 \\
23  &  7.3  & 2.2  & 1.3  & 0.9  & 0.5  & 0.4  & 0.4  & 0.3  & 0.1  & 0.01  &11.3  \\
24  & 11.9  & 3.6  & 2.1  & 1.4  & 0.8  & 0.7  & 0.6  & 0.6  & 0.2  & 0.02  &22.5 \\
25  & 10.3  & 3.1  & 1.8  & 1.2  & 0.7  & 0.6  & 0.5  & 0.5  & 0.1  & 0.02  &18.7 \\
\enddata
\tablenotetext{a}{First ten columns, A(V) to A(M2), refer to the extinction values 
derived using the A(V) from Schlegel et al.\ (1998)}
\end{deluxetable}

\begin{deluxetable}{rrrrrrrrrr}
\tablecaption{Extinction-Corrected\tablenotemark{a} Aperture Flux Densities in MilliJansky of Spitzer Galaxies in the ZoA \label{tbl:phot}}
\tablewidth{0pt}
\tablecolumns{10}
\tablehead{ \colhead{ID} &\colhead{J1.25} &\colhead{H1.65} &\colhead{Ks2.17} &\colhead{IRAC3.6} &\colhead{IRAC4.5} &\colhead{IRAC5.8} &\colhead{IRAC8} &\colhead{MIPS24} &\colhead{MIPS70} }
\startdata
1   & 6.9    & 6.7   & 2.1  & 3.530  & 2.120  & 3.275  & 7.109   & \nodata\tablenotemark{b} & \nodata\tablenotemark{b} \\
2   & 22.5   & 20.7  & 15.7 & 9.315  & 6.236  & 19.397 & 14.117  & 45.9    & \nodata \\
3   & 155.0  & 55.2  & 41.7 & 22.848 & 17.238 & 40.437 & 106.756 & 88.4    & 791.7   \\
4   & 848.0  & 117.0 & 30.1 & 9.167  & 5.866  & 5.907  & 28.262  & 9.9     & \nodata \\
5   & 339.0  & 81.7  & 24.9 & 10.866 & 8.078  & 21.896 & 71.899  & 35.7    & \nodata \\
6   & 401.0  & 87.6  & 29.2 & 10.316 & 7.196  & 9.225  & 25.140  & 11.9    & \nodata \\
7   & 3290.0 & 349.0 & 91.2 & 16.589 & 10.096 & 2.152  & 15.965  & 8.5     & \nodata \\
8   & 38.0   & 31.4  & 21.2 & 12.202 & 8.212  & 20.278 & 56.766  & 49.3    & 663.0   \\
9   & 72.5   & 28.7  & 11.9 & 6.306  & 4.060  & 4.154  & 16.271  & 9.4     & \nodata \\
10  & 754.0  & 124.0 & 46.0 & 11.660 & 8.217  & 8.161  & 29.893  & 8.5     & 179.4   \\
11  & 34.6   & 16.3  & 3.7  & 3.484  & 2.014  & 4.197  & 6.766   & 3.4     & \nodata \\
12  & 1840.0 & 138.0 & 31.6 & 13.225 & 8.223  & 8.541  & 33.679  & 10.2    & 156.0   \\
13  & 16.7   & 5.0   & 3.0  & 2.761  & 1.917  & 0.534  & 7.889   & \nodata\tablenotemark{b} & \nodata\tablenotemark{b} \\
14  & 545.0  & 105.0 & 28.2 & 14.223 & 9.692  & 14.856 & 37.765  & 25.5    & 280.8   \\
15  & 14.9   & 2.9   & 4.6  & 2.926  & 1.946  & 2.379  & 9.957   & 5.1     & 62.4    \\
16  & 59.6   & 33.5  & 35.8 & 9.162  & 6.077  & 5.782  & 28.286  & 15.6    & 89.7    \\
17  & 25.6   & 12.1  & 7.9  & 3.793  & 2.826  & 3.832  & 17.482  & \nodata\tablenotemark{b} & \nodata\tablenotemark{b} \\
18  & 26.9   & 16.2  & 11.2 & 3.997  & 2.931  & 1.145  & 11.450  & 3.7     & \nodata \\
19  & 16.5   & 12.6  & 9.2  & 6.156  & 4.527  & 7.414  & 27.557  & 16.2    & 328.0   \\
20  & 99.9   & 27.6  & 11.9 & 3.892  & 3.471  & 6.628  & 19.335  & 8.8     & 216.1   \\
21  & 7.9    & 5.2   & 2.6  & 2.435  & 0.885  & 2.276  & 5.383   & 4.3     & \nodata \\
22  & 14.1   & 8.8   & 6.0  & 3.976  & 2.383  & 2.308  & 8.040   & 5.5     & \nodata \\
23  & 8.2    & 6.7   & 6.9  & 3.912  & 2.917  & 3.508  & 10.329  & 5.4     & \nodata \\
24  & 32.8   & 14.6  & 5.3  & 4.899  & 2.948  & 5.066  & 13.529  & 6.4     & \nodata \\
25  & 32.6   & 14.4  & 5.1  & 4.700  & 2.700  & 4.800  & 13.333  & 6.3     & \nodata \\
\enddata
\tablenotetext{a}{Based on Schlegel et al.\ (1998)}
\tablenotetext{b}{Outside of MIPS survey}
\end{deluxetable}

\begin{deluxetable}{rrrrrrrrrr}
\tablecaption{Extinction-Corrected\tablenotemark{a} Aperture Flux Densities in MilliJansky of Spitzer Galaxies in the ZoA \label{tbl:photsed}}
\tablewidth{0pt}
\tablecolumns{10}
\tablehead{ \colhead{ID} &\colhead{J1.25} &\colhead{H1.65} &\colhead{Ks2.17} &\colhead{IRAC3.6} &\colhead{IRAC4.5} &\colhead{IRAC5.8} &\colhead{IRAC8} &\colhead{MIPS24} &\colhead{MIPS70} }
\startdata
 1  & 2265.7  &  214.2  &   20.4  &   12.576  &    6.230  &    8.411  &   17.233  & \nodata\tablenotemark{b} & \nodata\tablenotemark{b} \\
 2  &   53.1  &   34.6  &   22.0  &   11.247  &    7.317  &   22.310  &   16.098  &   47.7  & \nodata \\
 3  &  409.0  &   98.6  &   61.0  &   28.265  &   20.648  &   47.356  &  123.820  &   92.3  &  796.3 \\
 4  &  390.2  &   73.6  &   22.2  &    7.732  &    5.077  &    5.206  &   25.101  &    9.6  & \nodata \\
 5  &  320.7  &   79.0  &   24.4  &   10.735  &    7.995  &   21.699  &   71.292  &   35.6  & \nodata \\
 6  &  250.3  &   66.1  &   24.3  &    9.303  &    6.592  &    8.544  &   23.393  &   11.7  & \nodata \\
 7  &   55.9  &   30.5  &   18.5  &    6.788  &    4.730  &    1.108  &    8.564  &    7.1  & \nodata \\
 8  &   68.0  &   44.5  &   26.6  &   13.863  &    9.151  &   22.294  &   62.048  &   50.6  &  665.3 \\
 9  &  219.8  &   55.7  &   18.4  &    8.042  &    4.990  &    4.976  &   19.276  &    9.9  & \nodata \\
10  &  117.7  &   40.8  &   22.2  &    7.759  &    5.816  &    6.031  &   22.506  &    7.8  &  177.4 \\
11  &  653.6  &   94.5  &   11.7  &    6.636  &    3.479  &    6.772  &   10.602  &    3.9  & \nodata \\
12  & 1891.7  &  140.3  &   31.9  &   13.306  &    8.266  &    8.580  &   33.822  &   10.2  &  156.0 \\
13  &  404.9  &   33.6  &   10.5  &    5.555  &    3.469  &    0.897  &   12.842  & \nodata\tablenotemark{b} & \nodata\tablenotemark{b} \\
14  &  948.8  &  146.3  &   35.0  &   16.062  &   10.745  &   16.259  &   41.105  &   26.1  &  281.7 \\
15  &   61.3  &    6.8  &    8.0  &    3.989  &    2.532  &    2.995  &   12.359  &    5.4  &   62.9 \\
16  &   11.0  &   12.2  &   18.4  &    6.323  &    4.437  &    4.391  &   21.844  &   14.5  &   88.8 \\
17  &   42.2  &   16.3  &    9.6  &    4.232  &    3.101  &    4.156  &   18.867  & \nodata\tablenotemark{b} & \nodata\tablenotemark{b} \\
18  &   18.2  &   12.8  &    9.6  &    3.671  &    2.727  &    1.075  &   10.791  &    3.6  & \nodata \\
19  &   94.6  &   35.8  &   18.2  &    9.029  &    6.265  &    9.852  &   35.988  &   17.5  &  331.4 \\
20  &   73.6  &   23.0  &   10.6  &    3.640  &    3.280  &    6.307  &   18.455  &    8.7  &  215.7 \\
21  &   13.0  &    7.0  &    3.2  &    2.717  &    0.971  &    2.469  &    5.810  &    4.4  & \nodata \\
22  &   58.0  &   20.5  &   10.4  &    5.421  &    3.100  &    2.905  &    9.979  &    5.9  & \nodata \\
23  &   24.9  &   13.0  &   10.7  &    4.989  &    3.585  &    4.202  &   12.237  &    5.7  & \nodata \\
24  &  619.6  &   84.6  &   16.8  &    9.331  &    5.093  &    8.174  &   21.199  &    7.3  & \nodata \\
25  &  334.6  &   58.0  &   12.7  &    7.832  &    4.164  &    7.013  &   19.032  &    7.0  & \nodata \\
\enddata
\tablenotetext{a}{Based on our SED fits}
\tablenotetext{b}{Outside of MIPS survey}
\end{deluxetable}

\begin{deluxetable}{rc}
\tablecaption{Redshift Estimates of Spitzer Galaxies in the ZoA \label{tbl:z}}
\tablewidth{0pt}
\tablecolumns{2}
\tablehead{ \colhead{Name}  &\colhead{Redshift (SED fit)} }
\startdata
SPITZER191050+113409   &0.028 \\
SPITZER192252+124513   &0.026 \\
SPITZER192404+145632   &0.016 \\
SPITZER193206+183140   &0.026 \\
SPITZER193145+181940   &0.025 \\
SPITZER193133+181516   &0.025 \\
SPITZER193105+182614   &0.029 \\
SPITZER193337+190727   &0.024 \\
SPITZER193320+193334   &0.029 \\
SPITZER193349+202958   &0.027 \\
SPITZER193915+201426   &0.037 \\
SPITZER193825+212701   &0.022 \\
SPITZER193324+214918   &0.038 \\
SPITZER193956+213017   &0.021 \\
SPITZER194223+212514   &0.044 \\
SPITZER193633+225125   &0.029 \\
SPITZER193608+225054   &0.040 \\
SPITZER193632+225355   &0.040 \\
SPITZER193958+235251   &0.029 \\
SPITZER194042+232429   &0.038 \\
SPITZER194211+242507   &0.068 \\
SPITZER194257+251849   &0.038 \\
SPITZER194349+251449   &0.038 \\
SPITZER194919+243738   &0.031 \\
SPITZER195443+260745   &0.035 \\
\enddata
\end{deluxetable}

\end{document}